\documentclass[aps,preprint,preprintnumbers,amsmath,amssymb,floatfix,showpacs,nofootinbib]{revtex4}
\usepackage{graphicx}
\usepackage{epsfig}

\usepackage{amsbsy,bm}

\renewcommand{\H}{H$_2$~}
  
  \newcommand{\Hp}{H$_2^+$~}
\newcommand{\bxi}{\boldsymbol{\xi}}
\newcommand{\ba}{\begin{eqnarray}}
\newcommand{\ea}{\end{eqnarray}}
\newcommand{\br}{\begin{eqnarray*}}
\newcommand{\er}{\end{eqnarray*}}
\newcommand{\be}{\begin{equation}}
\newcommand{\ee}{\end{equation}}
\newcommand{\nn}{\nonumber}
\newcommand{\eref}[1] {(\ref{#1})}
\newcommand{\Eref}[1] {Eq.~(\ref{#1})}
\newcommand{\Fref}[1] {Fig. \ref{#1}}

\newcommand{\np}{\newpage}

\begin{document}

\title{Angular anisotropy of time delay in XUV/IR photoionization of \Hp}

\author{Vladislav V. Serov}
\affiliation{
Department of Theoretical Physics, Saratov State University, 83
Astrakhanskaya, Saratov 410012, Russia}

\author{A. S. Kheifets}

\affiliation{Research School of Physical Sciences,
The Australian National University,
Canberra ACT 0200, Australia}

\date{\today}
\begin{abstract}
We develop a novel technique for modeling of atomic and molecular
ionization in superposition of  XUV and IR fields with characteristics
typical for attosecond streaking and RABBITT experiments. The method is
based on solving the time-dependent Schr\"odinger equation in the
coordinate frame expanding along with the photoelectron wave
packet. The efficiency of the method is demonstrated by calculating
angular anisotropy of photoemission time delay of the H$_2^+$ ion in a
field configuration of recent RABBITT experiments.
\end{abstract}

\pacs{33.20.Xx, 33.80.Eh, 32.80.Fb}


\maketitle

\section{Introduction}

Attosecond time delay in laser induced photoemission of atoms and
molecules is a recently discovered phenomenon of ultrafast electron
dynamics. Following the pioneering experiments on two-color XUV/IR
photoionization \cite{M.Schultze06252010,PhysRevLett.106.143002},
various aspects of photoemission time delay have been thoroughly
investigated \cite{RevModPhys.87.765}.  One of such aspects is angular
anisotropy of the time delay relative to the joint polarization axis
of the XUV and IR light. Such an angular dependence is natural for
single XUV photon ionization of an $np$ atomic shell due to
interference of the $\epsilon s$ and $\epsilon d$ photoelectron
continua \cite{0953-4075-48-2-025602,Dahlstrom2014}. In two-color
XUV/IR photoionization, such an angular anisotropy can manifest itself
even in photoemission from a fully symmetric $ns$ atomic shell as it
has been demonstrated recently for the helium atom
\cite{2015arXiv150308966H}.  For more complex targets like molecules,
the angular dependence of the time delay brings particularly useful
information as it is sensitive to the orientation of the molecular
axis \cite{PhysRevA.87.063414}.

Because of low intensities of XUV and IR fields in a typical time
delay measurement, its theoretical modeling can be based on the lowest
order perturbation theory (LOPT) \cite{Dahlstrom2012}. More
punctilious approach requires an accurate solution of the
time-dependent Schr\"odinger equation for an atom or a molecule driven
by a combination of XUV and IR pulses as in an attosecond streaking
experiment, or an attosecond pulse train (APT) and an IR pulse in
RABBITT (Reconstruction of Attosecond Beating By Interference of
Two-photon Transitions). This solution can now be reliably obtained
for atomic targets with one or two active electrons
\cite{Whelan2013,PhysRevLett.113.263001}. However, due to the lack of
the spherical symmetry, the same solution becomes computationally
challenging for molecular targets.  To meet this challenge, we develop
a more efficient approach and seek a solution of the TDSE in a
coordinate frame which expands along with the photoelectron wave
packet \cite{PhysRevA.75.012715}. In addition, we employ a fast
spherical Bessel transformation (SBT) for the radial variables
\cite{2015arXiv150907115S}, a discrete variable representation for the
angular variables and a split-step technique for the time
evolution. This numerical approach allows us to reach space sizes and
propagation times hardly attainable by other techniques. Also, the use
of SBT ensures the correct phase of the wave function for a long time
evolution which is particularly important in time delay calculations.
To calibrate our technique, we reproduce the time delay values known from
the literature for the hydrogen \cite{Dahlstrom2012} and helium
\cite{2015arXiv150308966H} atoms. To demonstrate efficacy of our
numerical approach, we evaluate angular anisotropy of photoemission
time delay of the H$_2^+$ ion in a typical RABBITT experiment. Unlike
in atomic spherically symmetric targets, the angular anisotropy of
time delay in photoemission of \Hp is very strong due to interplay of
the two quantization axes: the polarization axis of light and the
interatomic molecular axis. The two aligned hydrogen nuclei act as a
double slit and cause a significant interference of the photoelectron
wave packet \cite{Akoury09112007,PhysRevA.81.062101}.  The
interference minima in the photoelectron spectra make their strong
imprint on the angular dependent part of the time delay.  The depth of
the minima increases close to the threshold where the normally
dominant dipole component of the ionization amplitude goes through its
Cooper minimum and give way to the octupole component.

\section{Method}

\subsection{The attosecond streaking}

We restrict ourselves with a single active electron (SAE) approximation and write the TDSE as
\begin{eqnarray}
i\frac{\partial\Psi(\mathbf{r},t)}{\partial
t}=\hat{H}\Psi(\mathbf{r},t)  
\label{TDSE}
\end{eqnarray}
with the Hamiltonian
\begin{eqnarray}
\hat{H}=\frac{\hat{p}^{\,2}}{2}-\mathbf{A}(t)\hat{\mathbf{p}}+U(\mathbf{r}).
\end{eqnarray}
Here $\hat{\mathbf{p}}=-i\nabla$ is the momentum operator,
$U(\mathbf{r},t)$ is the electron-nucleus interaction, $\mathbf{A}(t)$
is the vector potential of the electromagnetic field. The latter is
defined as 
\footnote{The atomic units are in use throughout the paper such that $e=m=\hbar=1$.  The factor $1/c$ with the speed of light $c\simeq 137$ and the electron charge $q=-1$ are absorbed into the vector potential.}
\begin{eqnarray}
\mathbf{A}(t) = -\int_0^t q \mathbf{E}(t')dt \ .
\end{eqnarray}
Here $\mathbf{E}(t)$ is the electric field vector.
In a typical attosecond streaking or a RABBITT experiment, the target
atom or molecule is exposed to a combination of the two fields:
\begin{eqnarray}
\mathbf{A}(t)= \mathbf{A}_{\rm XUV}(t) + \mathbf{A}_{\rm IR}(t-\tau)
\ ,
\end{eqnarray}
where $\tau$ is the relative displacement of the XUV and IR pulses.
We model an ultrashort XUV pulse by a Gaussian envelope
\begin{eqnarray}
\mathbf{A}_{\rm XUV}(t)=-\mathbf{n}_{\rm XUV}A_{\rm XUV}\exp\left(-2\ln 2
\frac{t^2}{\tau_{\rm XUV}^2}\right)\cos\omega_{\rm XUV} t \ ,
\end{eqnarray}
with the  FWHM $\tau_{\rm XUV}$. The IR pulse is described by the
$\cos^2$ envelope
\begin{eqnarray}
\mathbf{A}_{\rm IR}(t)=-\mathbf{n}_{\rm IR}A_{\rm IR} \cos^2(\pi t/\tau_{\rm IR})
\cos\omega t,\, |t|<\tau_{\rm IR}/2 \ ,
\end{eqnarray}
where $\tau_{\rm IR}$ is the IR pulse duration. The time evolution of
the target under consideration starts from the initial state
\begin{eqnarray}
\Psi(\mathbf{r},t_0)=\varphi_0(\mathbf{r})\exp(-iE_0t_0) \ , 
\label{Psi_init}
\end{eqnarray}
where $t_0=-\tau_{\rm IR}/2+\tau$ and  $\varphi_0(\mathbf{r})$,  $E_0$ 
are the wave function and the energy of the initial state.

After the end of the XUV pulse, the ionized electron is exposed to a
slow varying IR field and the long range Coulomb field of the residual
ion. The combination of these fields induces an additional correction
to the atomic time delay
\be
\tau_a = \tau_{\rm W} + \tau_{\rm CLC}
\ ,
\label{atomic}
\ee
where $\tau_{\rm W}$ is the Wigner time delay \cite{PhysRev.98.145}
and $\tau_{\rm CLC}$ is the Coulomb-laser coupling correction
\cite{0953-4075-44-8-081001}.  During the propagation in the IR field,
the photoelectron gains a considerable speed and travels large
distances from the parent ion. To describe this process, solution of
the TDSE should be sought in a very large coordinate box for a very
long propagation time which places a significant strain on
computational resources. To bypass this problem, we employ an
expanding coordinate system \cite{PhysRevA.75.012715}. In this method,
which we term the time-dependent scaling (TDS), the following variable
transformation is made:
\begin{eqnarray}
\mathbf{r} = a(t)\bxi \ . 
\label{r_xi}
\end{eqnarray}
Here  $a(t)$ is a scaling factor with an asymptotically linear time
dependence $a(t\to\infty)=\dot{a}_\infty t$ and $\bxi$ is a
coordinate vector. Such a transformation makes the coordinate frame to
expand along with the wave packet. In addition, the following
transformation is applied to the wave function
\begin{eqnarray}
\Psi(a(t)\bxi,t)=\frac{1}{[a(t)]^{3/2}}
\exp\left(\frac{i}{2}a(t)\dot{a}(t)\xi^2\right)\psi(\bxi,t). \label{Psi_psi}
\end{eqnarray}
Such a transformation removes a rapidly oscillating phase factor from
the wave function in the asymptotic region \cite{PhysRevA.75.012715}.  Thus
transformed wave function satisfies the equation
\begin{eqnarray}
i\frac{\partial\psi(\bxi,t)}{\partial
t}=\left[\frac{\hat{p}_\xi^{\,2}}{2[a(t)]^2}-\frac{\mathbf{A}(t)\hat{\mathbf{p}}_\xi}{a(t)}+U[a(t)\bxi]\right]\psi(\bxi,t)
\ ,
\label{TDSExi}
\end{eqnarray}
where 
$\hat{\mathbf{p}}_\xi=-i\nabla_\xi=
-i\left(\frac{\partial}{\partial\xi_x},\frac{\partial}
{\partial\xi_y},\frac{\partial}{\partial\xi_z}\right)$.
We note that if the spectrum of the operator $\hat{p}_\xi^{\,2}$ is
upper limited, which is the case for any numerical approximation of a
differential operator, then the first term in the RHS of \Eref{TDSExi}
tends to zero as $[a(t)]^{-2}$ for $t\to\infty$ . In the meantime, the
potential term with a long-range Coulomb asymptotic
$U(\mathbf{r}\to\infty)\sim 1/r$ is transformed to $U[a(t)\bxi]\sim
Z/a(t)\xi$. This means that both the Coulomb  term and the
vector potential term are decreasing in time as $1/a(t)$.
Therefore, when solving \Eref{TDSExi}, we can increase the time
propagation step $\Delta  t=a(t)\Delta t_0$ which accelerates 
the solution even further \cite{PhysRevA.75.012715}.

Remarkable property of the expanding coordinate system is that the
ionization amplitude $f(\mathbf{k})$ is related with the wave function
$\psi(\bxi,t)$ by a simple formula \cite{PhysRevA.75.012715}
\begin{eqnarray}
|f(\mathbf{k})|^2 = \dot{a}_\infty^{-3} \lim_{t\to\infty}
|\psi(\mathbf{k}/\dot{a}_\infty,t)|^2.
\end{eqnarray}
In practice, the evolution is traced for a very large time $t_f\gg
\tau_{\rm IR}$ and then the ionization probability density is obtained
from the expression 
\be P^{(3)}\equiv\frac{dP}{dk_xdk_ydk_z} = |f(\mathbf{k})|^2 \simeq
\dot{a}_\infty^{-3} |\psi(\mathbf{k}/\dot{a}_\infty,t_f)|^2.  \ee

The coordinate frame \eref{r_xi} is well suited for approximating an
expanding wave packet. However, its drawback is that the bound states
are described progressively less accurately as the coordinate frame
and its numerical grid expands. 
Therefore, during the XUV pulse, when an accurate approximation
of the bound states is required, we use a stationary coordinate
frame. The expansion of the frame starts at the moment $t_1\gg
\tau_{\rm XUV}$.
We use the piecewise linear scaling 
\begin{eqnarray}
a(t)=\left\{
\begin{array}{ll}
1,& t<t_1;\\
\dot{a}_\infty t,& t>t_1.
\end{array}
\right. \label{at}
\end{eqnarray}
At $t<t_1$ the wave function $\psi(\bxi,t)=\Psi(\mathbf{r},t)$.  Since
the time derivative of $a(t)$ defined by \Eref{at} have discontinuity at the
start of the expansion, the wave function at $t_1$ should be
multiplied by the phase factor
\begin{eqnarray}
\psi(\bxi,t_1+0)=\exp\left(\frac{i}{2}\dot{a}_\infty\xi^2\right)\psi(\bxi,t_1-0).
\end{eqnarray}
Here we choose $\dot{a}_\infty=1/t_1$. Such a choice ensures that the
wave packet remains stationary in the expanding frame at $t>t_1$.
To reduce  the initial state error from expanding frame, this state
is projected out from the wave packet by a simple orthogonalization
\begin{eqnarray}
\Psi(\mathbf{r},t_1)\to \Psi(\mathbf{r},t_1) - \left\langle
\varphi_0(\mathbf{r}) | \Psi(\mathbf{r},t_1) \right\rangle
\varphi_0(\mathbf{r}).
\end{eqnarray}
Other bound states are suppressed by introducing an imaginary
absorbing potential
\begin{eqnarray}
U_{sa}(\xi,t)=i\frac{\ln(1-e^{-\xi^2})}{a(t)}
\end{eqnarray}
This is equivalent to multiplying the wave function on
each step of the time propagation by the multiplier
$\exp(-iU_{sa}\Delta t)\approx (1-e^{-\xi^2})^{\Delta t/a(t)}$, which
tends to 0 at $\xi\to 0$. This way we introduce an absorbing mask with
the radius $\xi\sim 1$. As the coordinate frame expands, this mask
suppresses all the bound states but does not affect the expanding wave
function with the momenta $k \gg \dot{a}_\infty$.

\subsection{RABBITT}

In a RABBITT measurement, unlike in attosecond streaking, a target
atom or molecule is subjected to an attosecond pulse train (APT)
rather than a single XUV pulse. The APT field can be represented as
\begin{eqnarray}
\mathbf{A}_{\rm XUV}(t)=\sum_{\nu=-\left\lfloor N_{\rm
APT}/2\right\rfloor}^{\left\lfloor N_{\rm APT}/2\right\rfloor}(-1)^\nu
f_{\rm env}(t_\nu)\mathbf{A}^{(1)}_{\rm XUV}(t-t_\nu) \label{A_APT}
 \ ,
\end{eqnarray}
where $N_{\rm APT}$ is the number of pulses in the APT and  the arrival
time of each pulse
\begin{eqnarray}
t_\nu=\frac{T_{\rm IR}}{2}\nu,
\end{eqnarray}
is a half integer of the period of the IR oscillation $T_{\rm
 IR}=2\pi/\omega$.  The envelope of the APT is given by
\begin{eqnarray}
f_{\rm env}(t)=\exp\left(-2\ln 2 \frac{t^2}{\tau_{\rm APT}^2}\right)
\ ,
\end{eqnarray}
where $\tau_{\rm APT}$ is the FWHM. 

It is necessary to ensure an accurate representation of the bound
states during each of the pulses in the APT.  As the APT duration is
large, direct application of the expanding frame is not practical.
However, because the field intensity of the APT is usually small, we
can add contributions of each pulse to the ionized electron wave
packet by a simple summation. 

Let us coincide a set of the wave functions $\psi_\nu(\bxi,t)$
satisfying the equation
\begin{eqnarray}
i\frac{\partial\psi_\nu(\bxi,t)}{\partial
t}=\left[\frac{\hat{p}_\xi^{\,2}}{2a(t)^2}-\frac{\mathbf{A}_\nu(t)\hat{\mathbf{p}}_\xi}{a(t)}+U[a(t)\bxi]\right]\psi_\nu(\bxi,t). \label{TDSE_nu}
\end{eqnarray}
where
\begin{eqnarray}
\mathbf{A}_\nu(t)= \mathbf{A}_{\rm XUV}^{(1)}(t) + \mathbf{A}_{\rm IR}(t+t_\nu-\tau).
\end{eqnarray}
By taking into account the coordinate and momentum relation
 $\bxi\simeq\mathbf{k}/\dot{a}_\infty$ at large $t_f$, the APT perturbation
 of the wave function, orthogonalized to the ground state, can be
 expressed as 
\begin{eqnarray}
\psi(\bxi,t_f)=\sum_{\nu=-\left\lfloor N_{\rm
APT}/2\right\rfloor}^{\left\lfloor N_{\rm APT}/2\right\rfloor}(-1)^\nu
f_{\rm env}(t_\nu)\exp\left(i\frac{\dot{a}\xi^2}{2}
t_\nu\right)\psi_\nu(\bxi,t_f). \label{sum_psi}
\end{eqnarray}
The intensity of the IR pulse should be fairly large to ensure
sufficient intensity of the two-photon transitions. If such an IR
pulse is applied suddenly to the target before arrival of the APT,
this may cause a considerable unphysical distortion of the initial
state. To avoid this artifact, we applied the following initial
condition
\begin{eqnarray}
\psi_\nu(\mathbf{r},t_0)=\Psi_{\rm
IR}(\mathbf{r},t_0+t_\nu-\tau). \label{psi_nu_init}
\end{eqnarray}
Here $t_0<0$, $|t_0|\gg \tau_{\rm XUV}$, and the wave function
$\Psi_{\rm IR}(\mathbf{r},t)$ is a solution of the equation
\begin{eqnarray}
i\frac{\partial\Psi_{\rm IR}(\mathbf{r},t)}{\partial
t}=\left[\frac{\hat{p}^{\,2}}{2}-\mathbf{A}_{\rm IR}(t)\hat{\mathbf{p}}+U(\mathbf{r})\right]\Psi_{\rm IR}(\mathbf{r},t)
\label{TDSE_IR}
\end{eqnarray}
with the initial condition \eqref{Psi_init} that describes the
evolution in the IR field alone. As the low frequency IR field does
not cause a considerable ionization, such a solution does not expand
to large distances and can be modeled with a modest size of the
radial box.
 
In our approach, the resulting photoelectron spectrum is a simple sum
of the spectra induced by each of the $N_{\rm APT}$ pulses.  In the
case when $\tau_{\rm IR}\gg \tau_{\rm APT}$, the amplitude of the IR
field oscillation during ionization can be considered constant. Thus
the photoelectron spectrum can be constructed from just the two XUV
pulses of the opposite polarity overlapping with a single IR
oscillation. The remaining pulses are translated by an integer number
of IR periods.
According to the Floquet theory, the initial state wave function
satisfies the following periodic condition
\begin{eqnarray}
\Psi_{\rm IR}(\mathbf{r},t + n T_{\rm IR})=\Psi_{\rm
IR}(\mathbf{r},t)\exp(-iE_{Q} n T_{\rm IR}) \ , 
\label{FloquetPsi}
\end{eqnarray}
where $E_{Q}$ is the quasienergy. Hence
\begin{eqnarray}
\psi_{\nu+2n}(\bxi,t_f)=\psi_\nu(\bxi,t_f)\exp(-iE_{Q} n
T_{\rm IR}).\label{shift_psi}
\end{eqnarray}
Thus, by solving \Eref{TDSE_nu} and calculating $\psi_\nu(\bxi,t_f)$
for $\nu=0$ and $\nu=1$ with the initial condition \eref{psi_nu_init},
\Eref{shift_psi} allows to express all the other terms for
evaluating the sum in \Eref{sum_psi}.

A separate task is to evaluate the function $\Psi_{\rm
IR}(\mathbf{r},t)$ satisfying the periodic condition \eref{FloquetPsi}
and find the quasienergy $E_{Q}$.  This can be done by a direct solution of
\Eref{TDSE_IR} with the condition \eref{FloquetPsi}, or by the Floquet
series expansion. We, however, found a simpler way. We determined
the time evolution of $\Psi_{\rm IR}(\mathbf{r},t)$ with the initial
condition
$\Psi(\mathbf{r},t_{0{\rm IR}})=\varphi_0(\mathbf{r})\exp(-iE_0t_{0{\rm IR}})$
after the IR field iss gradually switched on
\begin{eqnarray}
\mathbf{A}'_{\rm IR}(t)=-\mathbf{n}_{\rm IR}A_{\rm IR}\cos\omega t
\times \left\{
\begin{array}{ll}
\exp\left(\displaystyle -\frac{(t-t_{\rm on})^4}{\tau_{\rm on}^4}\right),&
t<t_{\rm on};\\ 1, & t>t_{\rm on}.
\end{array}
\right.
\end{eqnarray}
Adiabatic switching and a smooth transition to the constant IR field
regime ensures that the wave function at $t>t_{\rm on}$ is close to the
true periodic solution. We used the following switching parameters
$t_{\rm on}=-0.75T_{\rm IR}$, $\tau_{\rm on}=T_{\rm IR}$ and started
the time evolution from $t_{0{\rm IR}}=-3\tau_{\rm on}+t_{\rm on}$.  The
quasienergy was extracted by projecting thus obtained function at the
end of the period onto the one determined at the beginning of the
period:
\begin{eqnarray}
E_{Q}=E_0-\rm{Im}
\left\{
\ln [\left\langle \Psi_{\rm IR}(\mathbf{r},
-T_{\rm IR}/2)| \Psi_{\rm IR}(\mathbf{r},T_{\rm IR}/2)
 \right\rangle e^{iE_0T_{\rm IR}}]\right\}/T_{\rm IR}. 
\end{eqnarray}
At the field intensity employed in our calculations, 
the quasienergy $E_{Q}$ differs from the ground state energy $E_0$
only in the fourth significant figure.

Because  \Eref{sum_psi} was derived under assumption of vanishing
external field at $t_f$,  $\psi_\nu(\bxi,t)$ was evaluated with a
smooth switching of the IR vector potential
\begin{eqnarray}
\mathbf{A}_{\rm IR}(t)=-\mathbf{n}_{\rm IR}A_{\rm IR}\cos\omega t \times
\left\{
\begin{array}{ll}
1, & t<t_{\rm off};\\
\exp\left(\displaystyle
-\frac{(t-t_{\rm off})^4}{\tau_{\rm off}^4}\right),& t>t_{\rm off}.
\end{array}
\right.
\end{eqnarray}
%
%
Here the switching time $t_{\rm off}$ and duration $\tau_{\rm off}$
were chosen very large, $t_{\rm off}=32T_{\rm IR}$, $\tau_{\rm
off}=4T_{\rm IR}$. The end of propagation was set to $t_f=t_{\rm
off}+5\tau_{\rm off}\approx 6000$.

\section{Results}

We solve Eqs.~\eref{TDSE_nu} and \eref{TDSE_IR} using a fast SBT
\cite{2015arXiv150907115S} for the radial variables, a discrete
variable representation for the angular variables and a split-step
technique for the time evolution.
In all the calculations, we set the box size to $\xi_{max}=51.2$~a.u. The
radial grid step was set to $\Delta \xi=0.1$~a.u. unless specified
differently. For atomic calculations on H and He, the angular basis
was restricted to $N_\theta=4$ spherical harmonics whereas for H$_2^+$
we used $N_\theta=16$.

The APT is modeled by a series of $N_{\rm APT}=11$ Gaussian pulses
with the width $\tau_{\rm XUV}=5$ a.u. (120 as) and the APT width
$\tau_{\rm APT}=2T_{\rm IR}$ (5.2~fs), whereas a long IR pulse is
modeled by a continuous wave with the frequency $\omega=0.05841$
a.u. (photon energy 1.59~eV, $\lambda=780$~nm) and the vector
potential amplitude $A_{\rm IR}=0.05$. The latter corresponds to the
electric field strength $\mathcal{E}_{\rm IR}=1.5\times 10^{9}$~V/m
and the field intensity $3\times 10^{11}$~W/cm$^2$.  The amplitude of
the XUV pulse was $A_{\rm XUV}=0.025$~a.u. (the field intensity $0.75\times
10^{11}(\omega_{\rm XUV}/\omega_{\rm IR})^2$~W/cm$^2$) The relative
APT/IR time delay $\tau$ was varied from 0 to 0.5$T_{\rm IR}$ with a
step 0.03125$T_{\rm IR}$.
By exposing an atom or a molecule to the APT \eref{A_APT} with the
central frequency $\omega_{\rm XUV}=(2q_{0}+1)\omega$, the photoelectrons
will be emitted with the energies
%
%
$E_{2q+1}=(2q+1)\omega-E_0$ 
corresponding to the odd harmonics of the IR frequency $\omega$. The heights of the
corresponding peaks will be Gaussian distributed with the center at
$E_{2q_0+1}$ and the width inversely proportional to the width of the
XUV pulse $\tau_{\rm XUV}$. The width of the individual photoelectron
peaks will be inversely proportional to the APT width $\tau_{\rm
APT}$. Superimposing a dressing IR field will add additional peaks in
the photoelectron spectrum at 
%
%
$E_{2q}=2q\omega-E_0 \ .$
These additional peaks, known as the sidebands (SB), correspond to the
even harmonics. The sideband amplitudes will vary with the relative
time delay $\tau$ of the APT and the IR pulses as \cite{Paul01062001}
\begin{eqnarray}
S_{2q}(\tau)=A+B\cos[2\omega(\tau-\tau_a)], \label{SBtau}
\end{eqnarray}
where $\tau_a$ is the atomic time delay \eref{atomic}. Here we assume
that there is no group delay (chirp) in the APT spectrum and all the harmonics
have the same phase.

\begin{figure}[ht]
	\centering
	\includegraphics[width=0.35\textwidth,angle=-90]{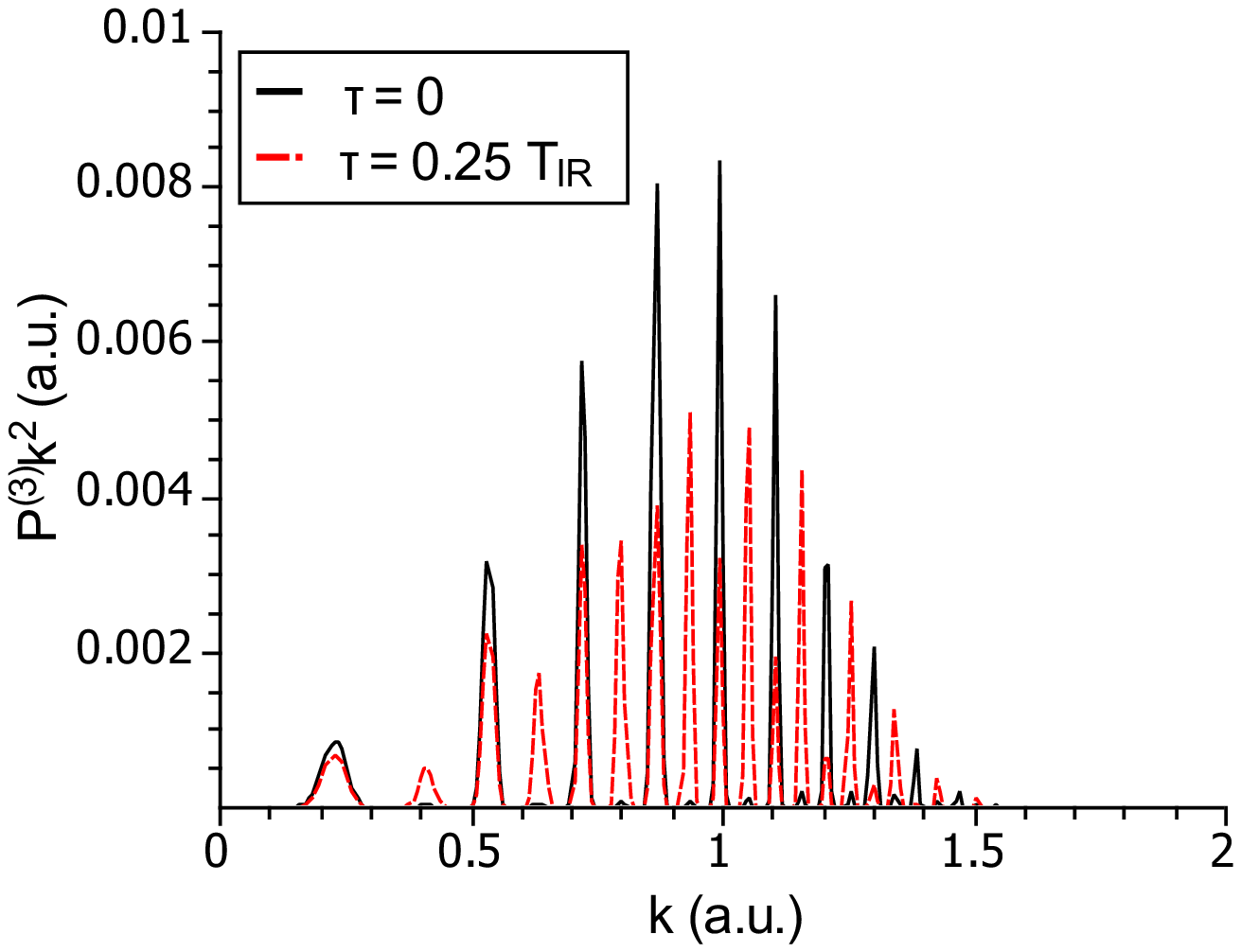}
	\includegraphics[width=0.35\textwidth,angle=-90]{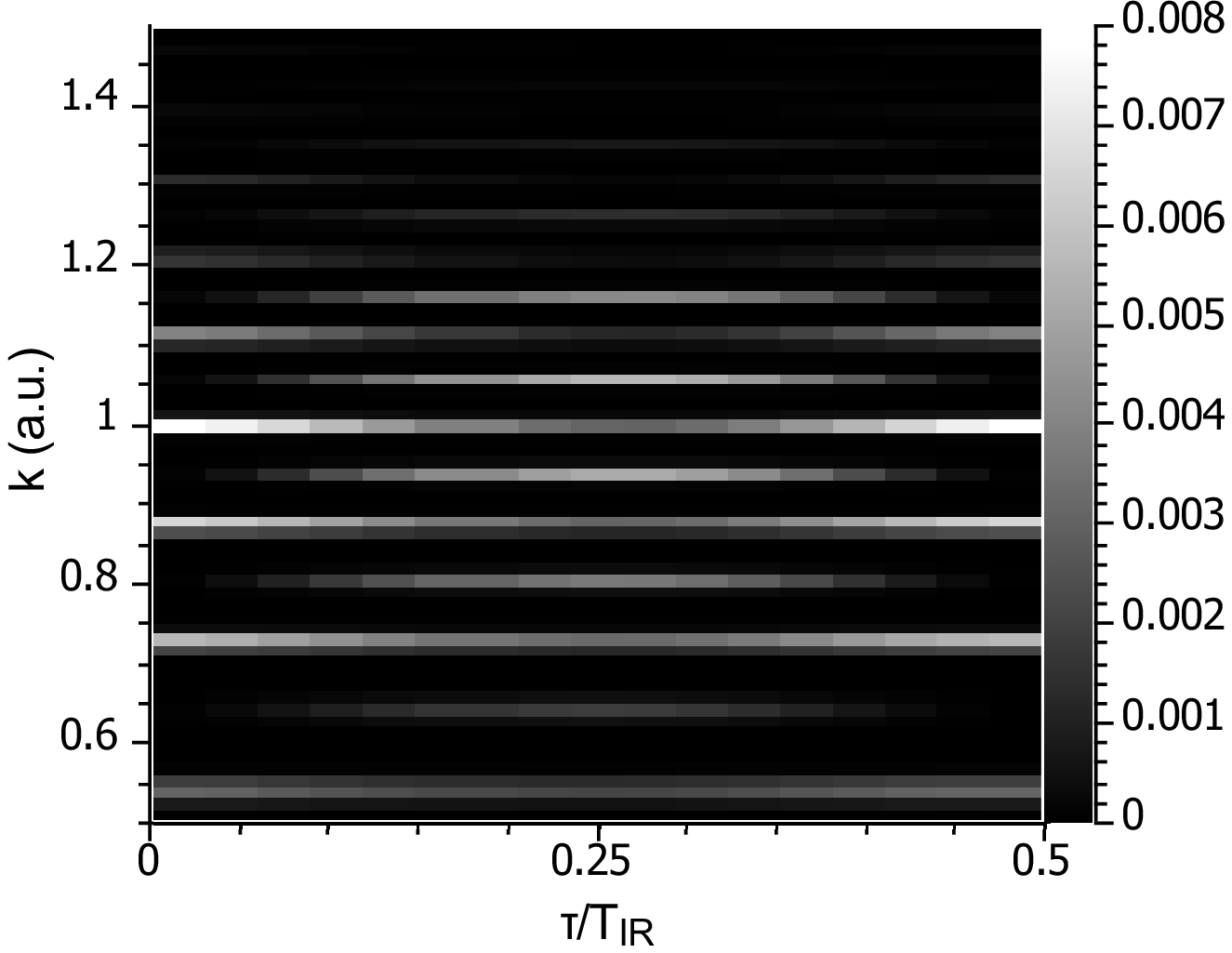}
	\caption{Photoelectron spectrum of the hydrogen atom in the
	polarization axis direction.  Left: the probability density
	$P^{(3)}k^2$ as a function of the photoelectron momentum $k$
	for two fixed values of $\tau$. Right: a gray scale map of
	$P^{(3)}(k,\tau)k^2$.}
	\label{fig:H_spectrum}
\end{figure}

\begin{figure}[ht]
	\centering
		\includegraphics[width=0.35\textwidth,angle=-90]{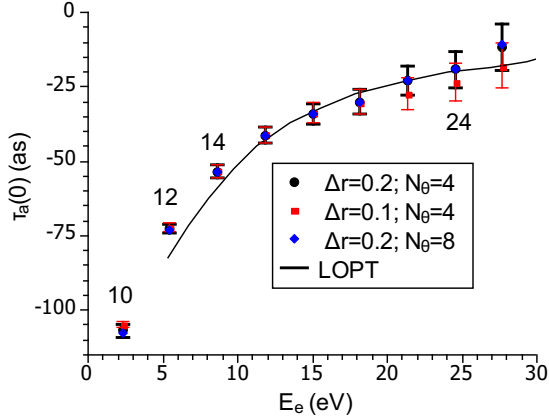}
	\caption{Atomic time delay $\tau_a$ of the hydrogen atom in
	the polarization axis direction as a function of the
	photoelectron energy $E_e$. Radial and angular numerical
	parameters are displayed in the legend. Error bars indicate
	the least squire fit uncertainty. The solid line visualizes
	the LOPT result of \citet{Dahlstrom2012}. Sideband indices
	made on the figure correspond to the four panels of
	\Fref{fig:H_RABITT_angle}.  }
	\label{fig:H_RABITT}
\end{figure}

\begin{figure}[ht]
	\centering
		\includegraphics[width=0.3\textwidth,angle=-90]{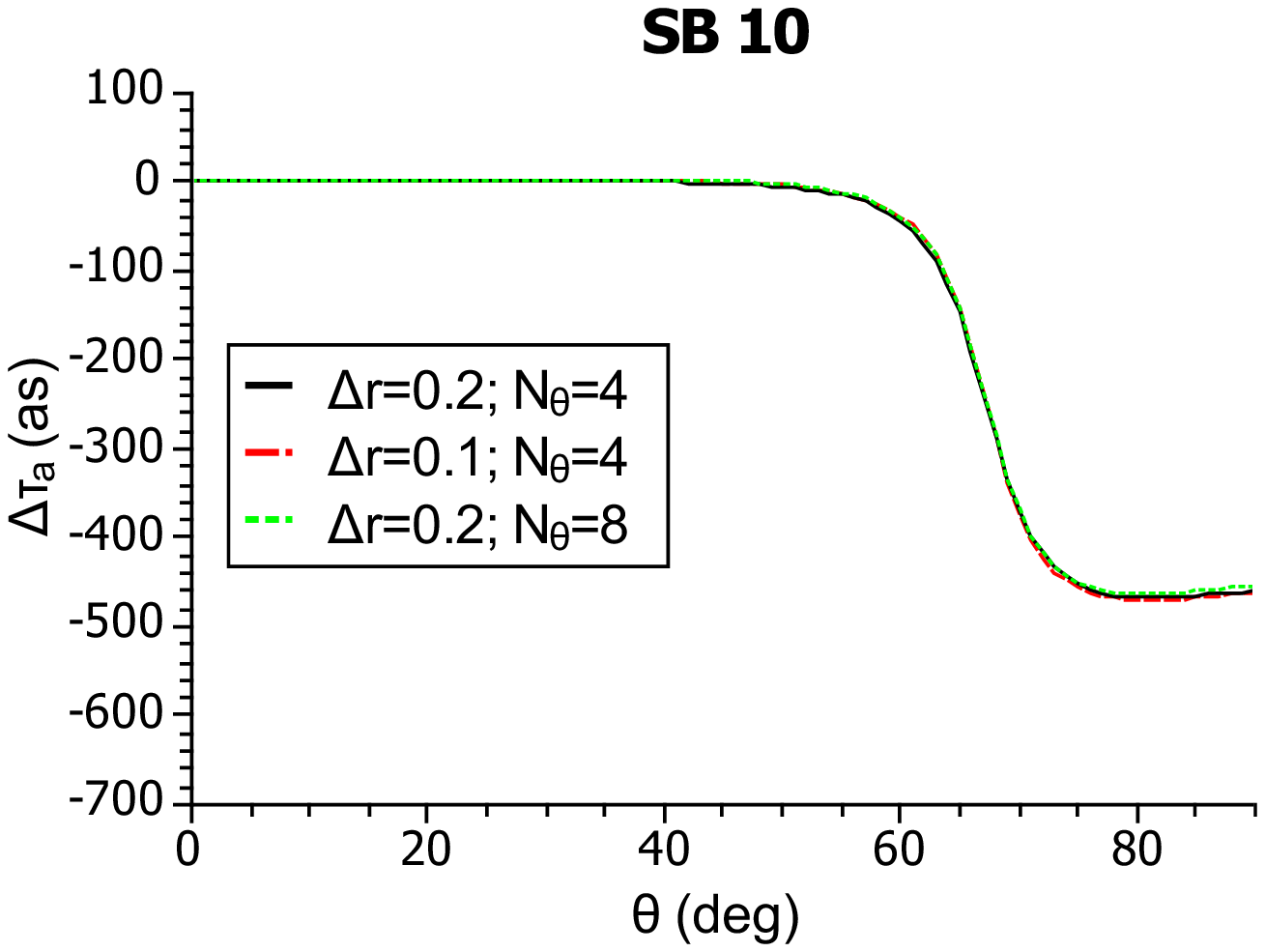}
		\includegraphics[width=0.3\textwidth,angle=-90]{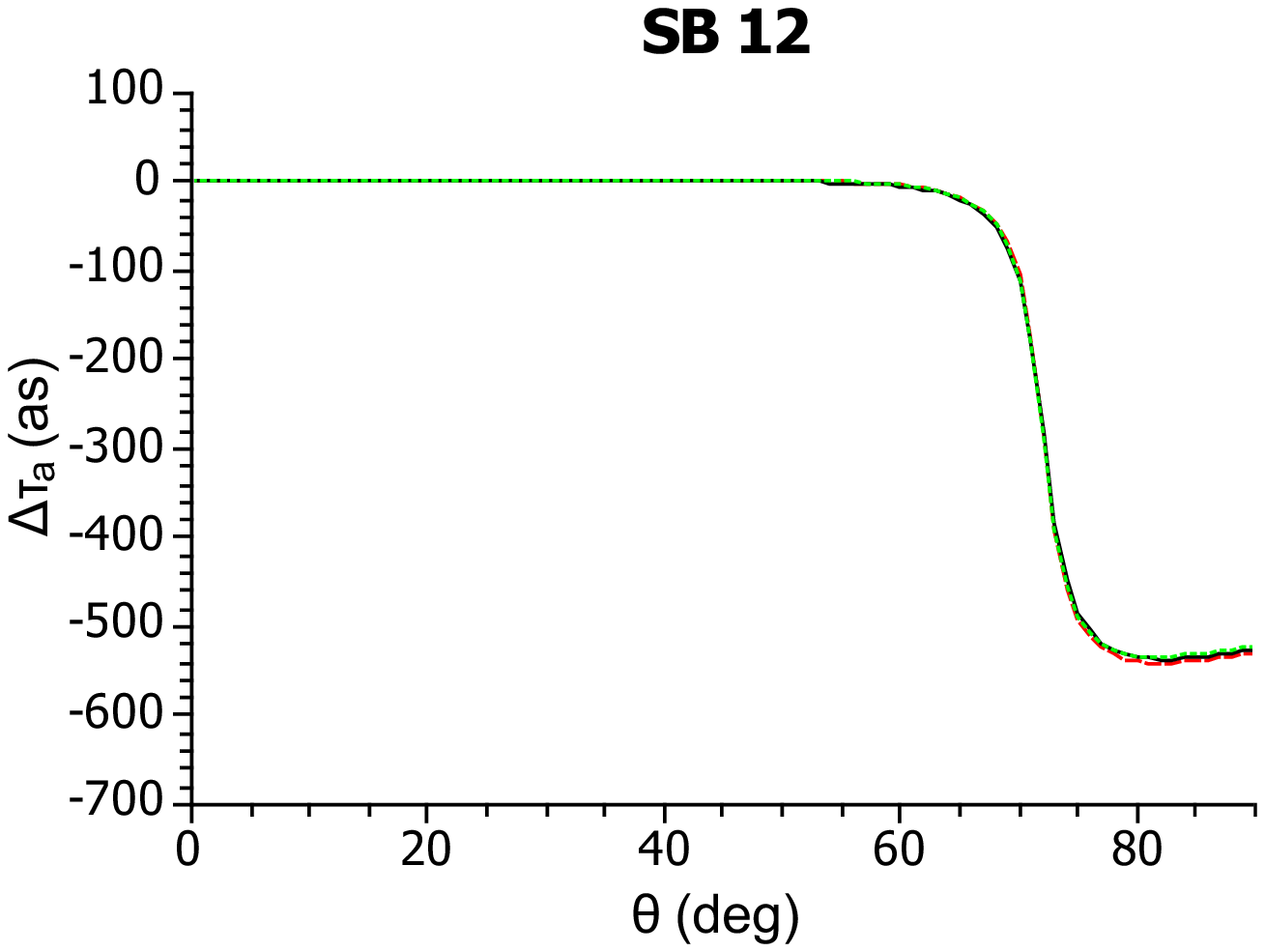}
		\includegraphics[width=0.3\textwidth,angle=-90]{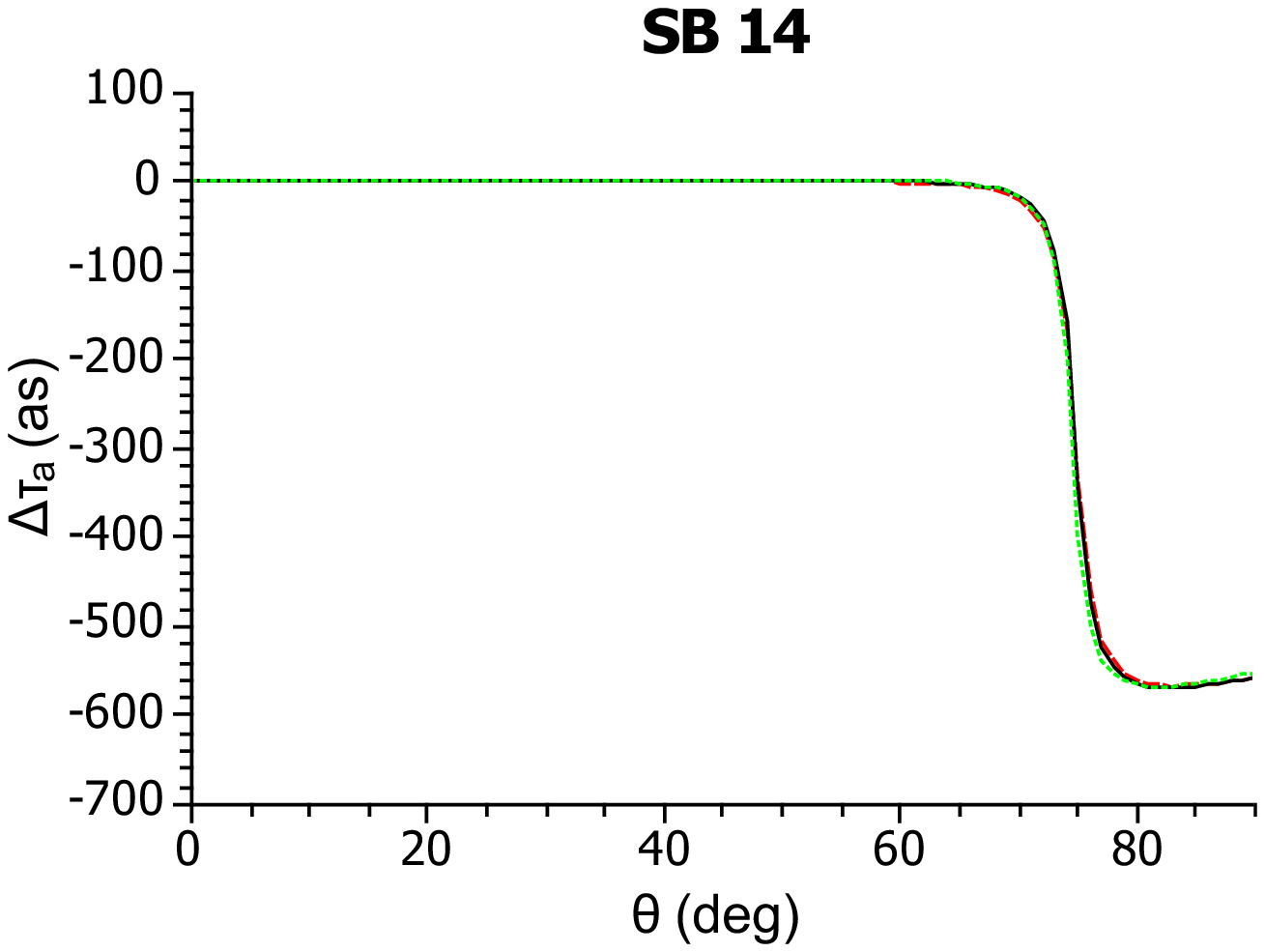}
		\includegraphics[width=0.3\textwidth,angle=-90]{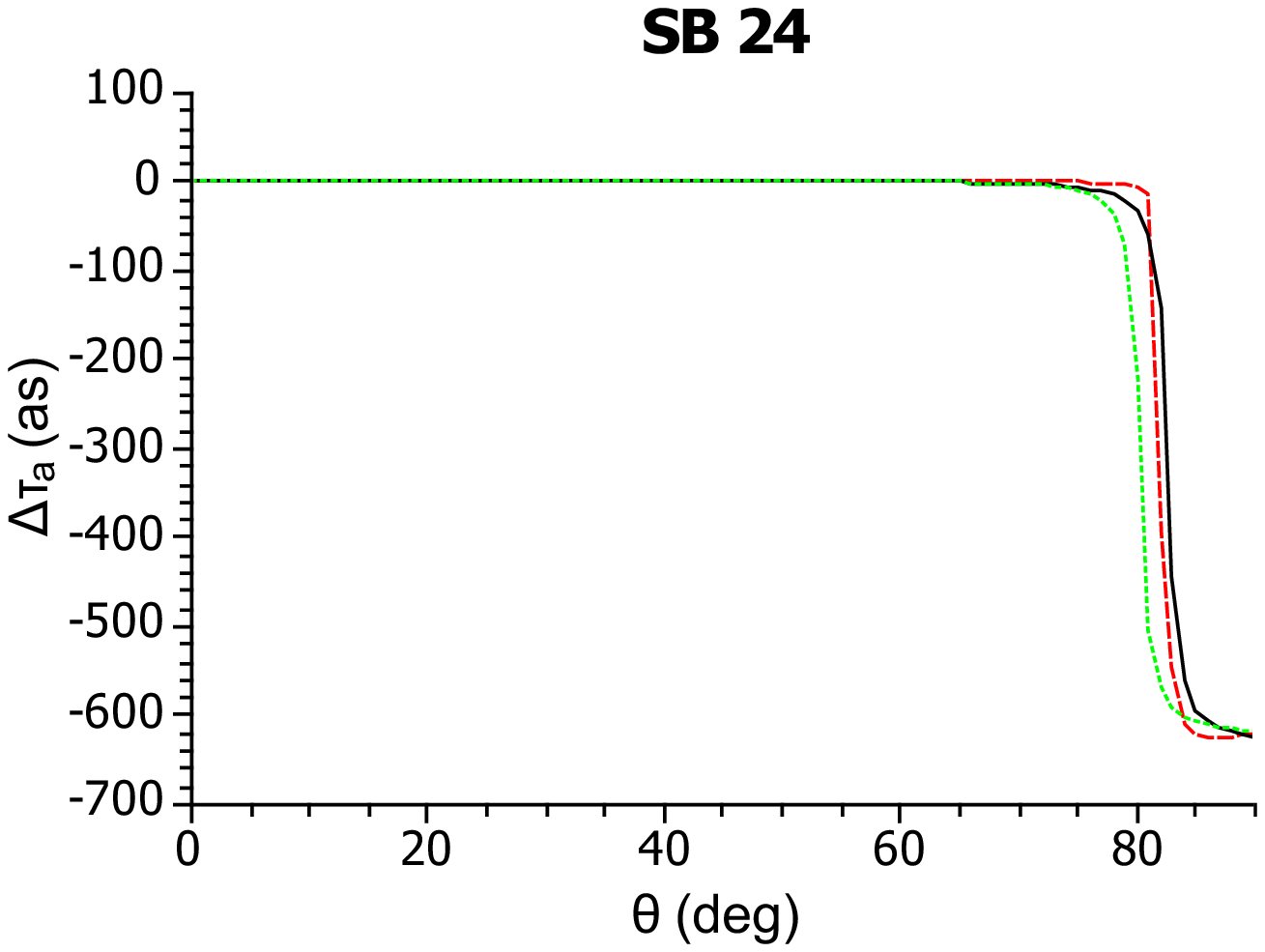}
	\caption{Angular-dependent part of the atomic time delay
	$\Delta\tau_a=\tau_a(E_e,\theta)-\tau_a(E_e,0)$ for the hydrogen atom as a
	function of the photoelectron emission angle $\theta$ relative
	to the polarization axis. The four panels correspond to
	different  sideband indices. Different line styles visualize three
	sets of radial and angular grid parameters as indicated in the
	legend of \Fref{fig:H_RABITT}.}
	\label{fig:H_RABITT_angle}
\end{figure}

This characteristic behavior is clearly seen in \Fref{fig:H_spectrum}
where we display the photoelectron spectrum of the hydrogen atom
subjected to an APT with the central frequency $\omega_{\rm
XUV}=17\omega$.  Here and in examples below, we set $\omega_{\rm XUV}$
such that the "central" peak in the photoelectron spectrum is
positioned at $E_{2q_0+1}\approx 0.5$~a.u.  We set the photoelectron
detection angle to $\theta=0$ which corresponds to the polarization
axis direction.  By the least square fit to \Eref{SBtau}, we obtained
the values of $\tau_a$ shown in \Fref{fig:H_RABITT}. Here the atomic
time delay is exhibited as a function of the photoelectron energy
$E_e=k^2/2$. The corresponding sideband indices are marked in the
figure. To test the numerical stability of our computational
procedure, we performed three sets of calculations: a) the radial grid
step $\Delta \xi=0.2$ and the number of spherical harmonics
$N_\theta=4$; b) $\Delta \xi=0.1$ and $N_\theta=4$; c) $\Delta
\xi=0.2$ and $N_\theta=8$. It is clearly seen from
\Fref{fig:H_RABITT_angle} that an increase of the angular basis size
$N_\theta$ does not affect the result. For lower photoelectron energy,
the time delay is not sensitive to the radial grid step. However, such
a sensitivity becomes noticeable for higher photoelectron energy
$E_e>25$~eV.

Same sensitivity to the radial and angular grid parameters can be seen
in Fig.\ref{fig:H_RABITT_angle} where we display the
angular dependent part of the atomic time delay
$\Delta\tau_a=\tau_a(E_e,\theta)-\tau_a(E_e,0)$ for the hydrogen atom as a function
of the photoelectron emission angle $\theta$ relative to the
polarization axis.  Based on this calibration, we restricted ourselves
to $\Delta \xi=0.1$ and $N_\theta=4$ to all the atomic calculations
shown below. For the H$_2^+$ ion, we used a larger angular basis
with $N_\theta=16$ to account the for the non-spherical ionic potential.

\begin{figure*}
	\centering
		\includegraphics[width=0.35\textwidth,angle=-90]{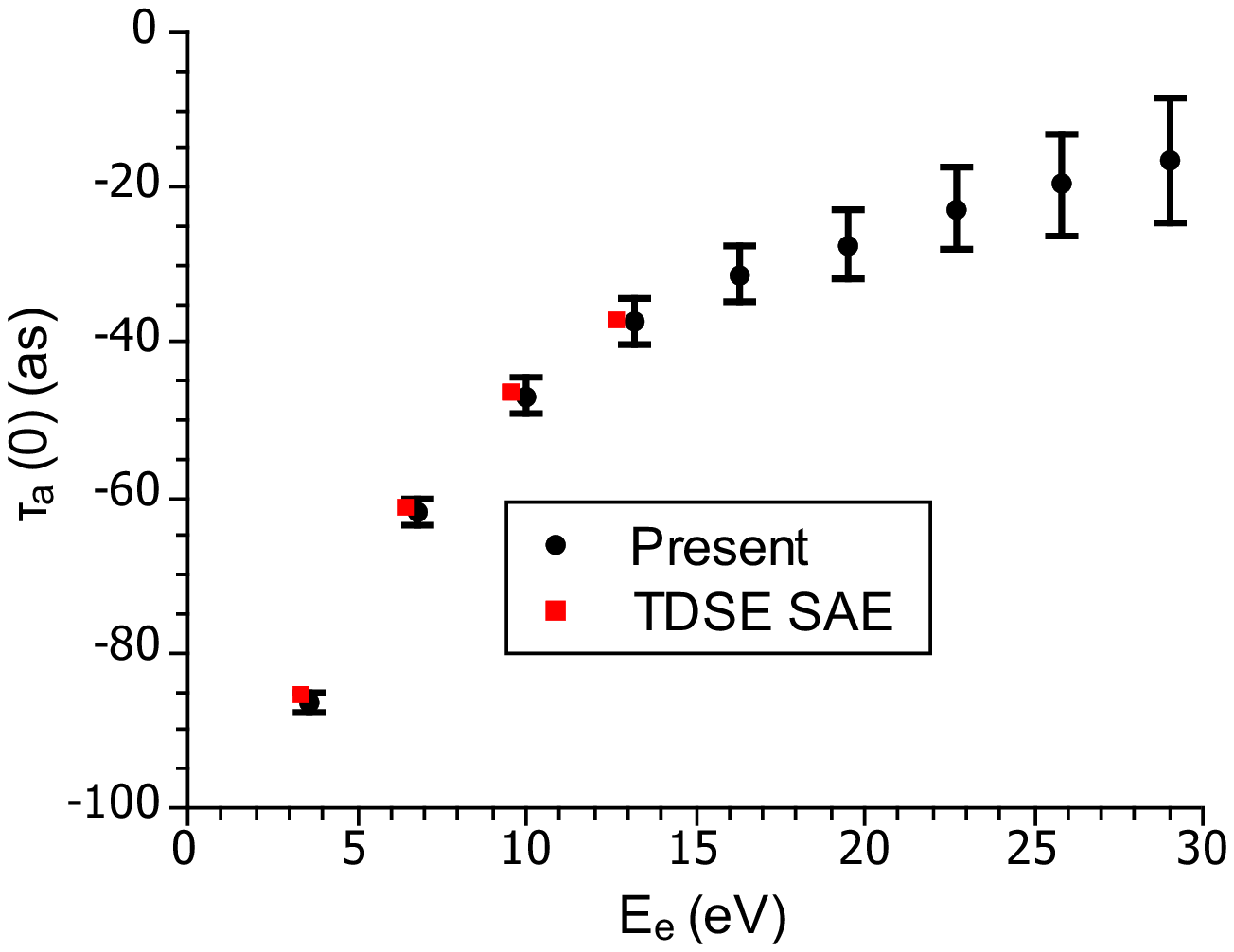}
	\caption{Atomic time delay $\tau_a$ of the helium atom as a
	function of $E_e$ for ejection angle $\theta=0^\circ$ and
	various sets of numerical parameters. Error bars display the
	uncertainty of the least square fit. The TDSE SAE results
	from \cite{2015arXiv150308966H} are also shown for comparison.}
	\label{fig:He_RABITT}
\end{figure*}

Further calibration of our technique is demonstrated in
\Fref{fig:He_RABITT} where we compare the atomic time delay of the
helium atom at $\omega_{\rm XUV}=25\omega$ with the results of direct
numerical solution of the TDSE in the SAE
\cite{2015arXiv150308966H}. In both sets of the TDSE calculations, the
non-local potential of the He atom was modeled by an analytical
parametrization \cite{Sarsa2004163}.  Close resemblance of the two
sets of data can be seen.

We note in passing that the numerical TDSE SAE results reported in
\cite{2015arXiv150308966H} required many hours of supercomputer time
whereas the present calculations were carried on a notebook computer in
less than an hour.

\begin{figure*}
	\centering
		\includegraphics[width=0.3\textwidth,angle=-90]{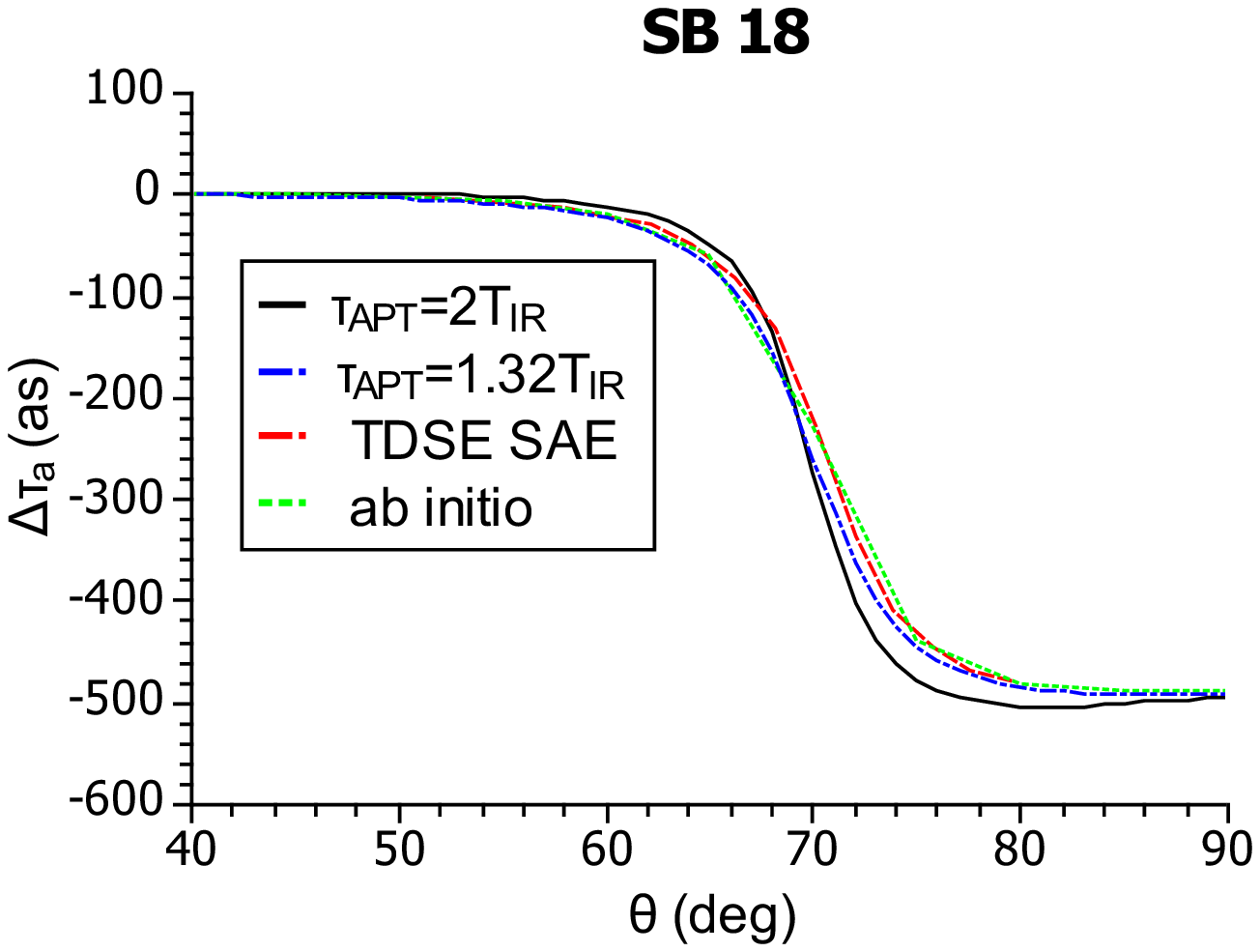}
		\includegraphics[width=0.3\textwidth,angle=-90]{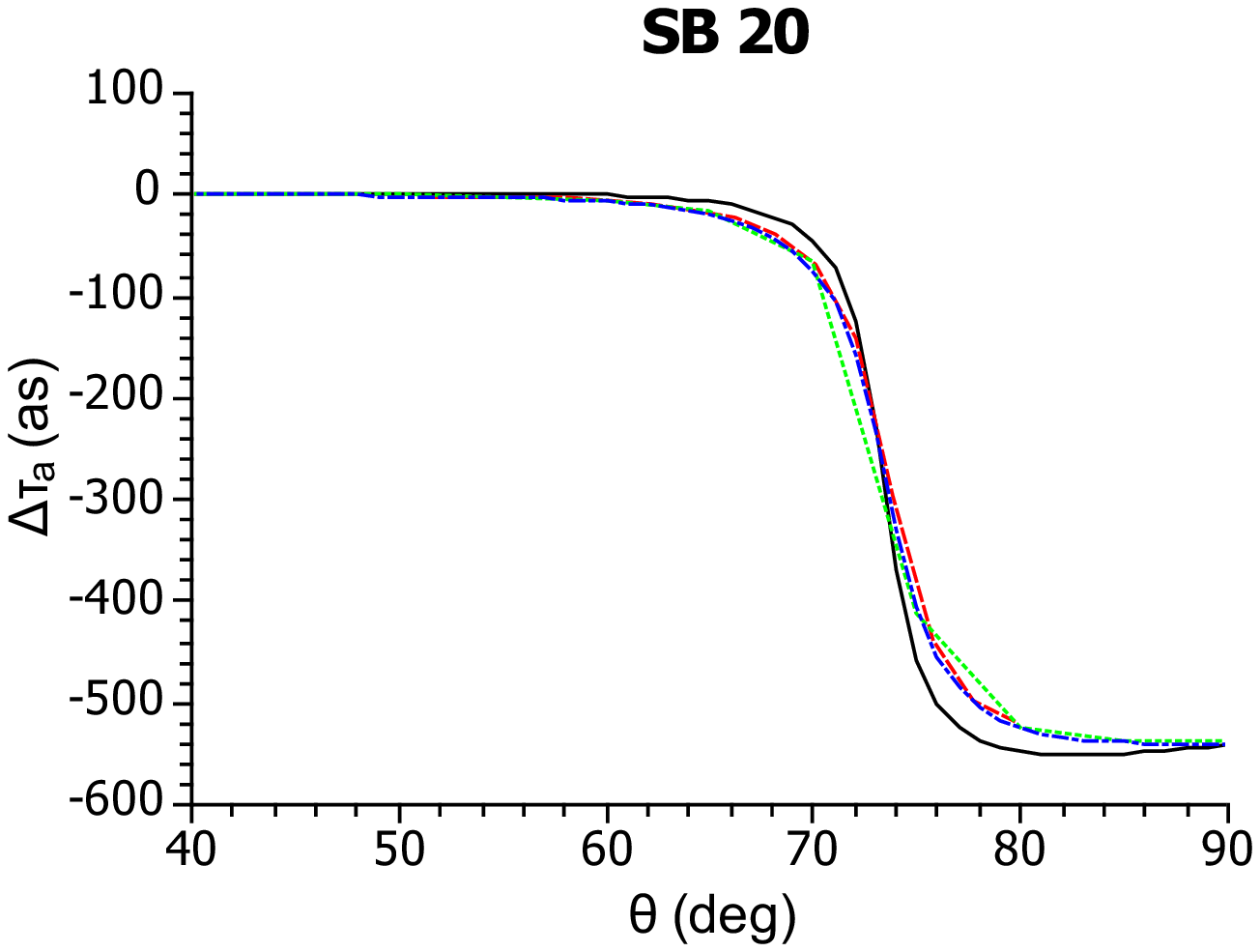}
		\includegraphics[width=0.3\textwidth,angle=-90]{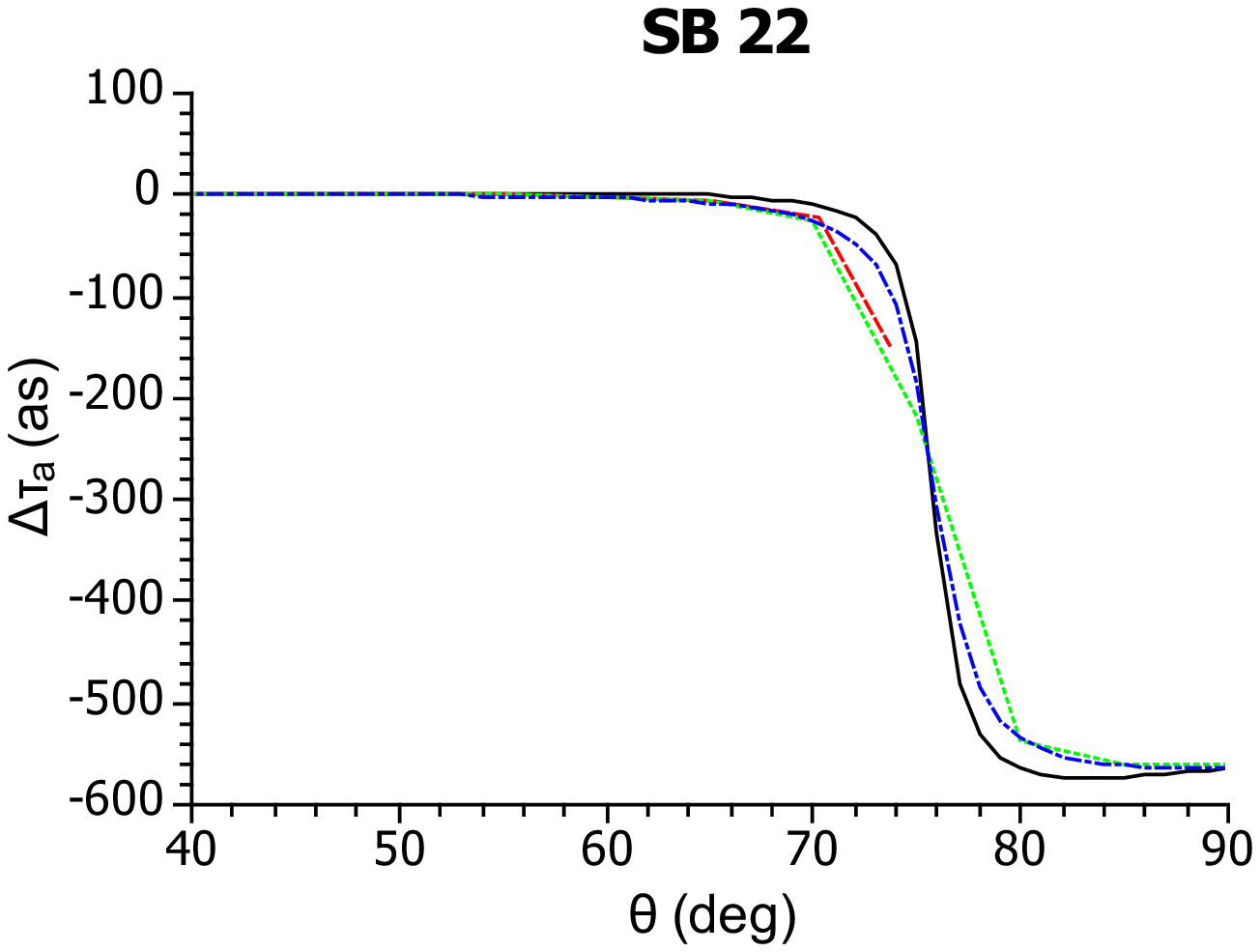}
		\includegraphics[width=0.3\textwidth,angle=-90]{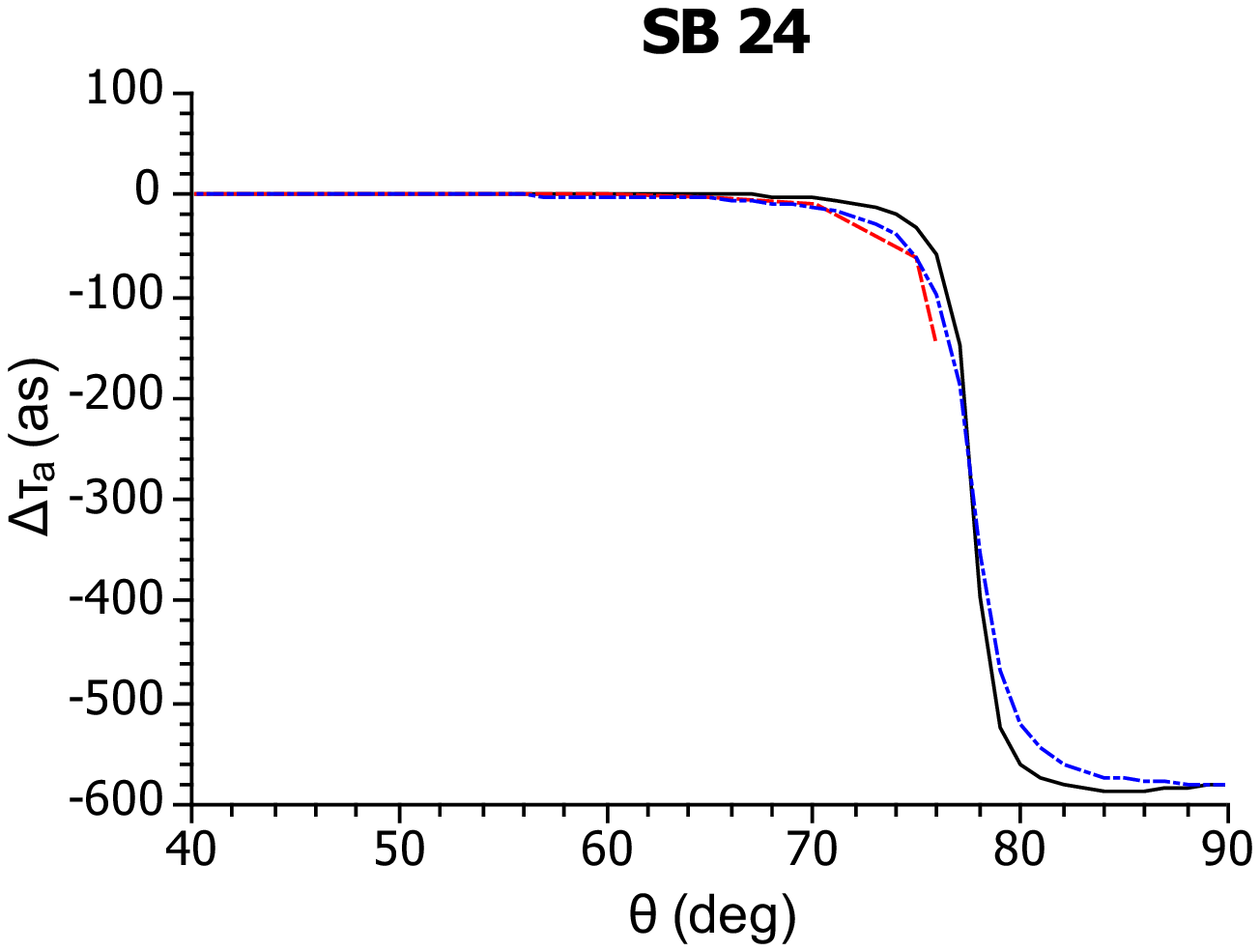}
	\caption{Angular-dependent part
	$\Delta\tau_a$ of the atomic time delay in
	He as function of electron ejection angle $\theta$ for four
	fixed SBs. The present results for $\tau_{\rm APT}=2T_{\rm IR}$
	(black solid line) and for $\tau_{\rm APT}=1.32T_{\rm IR}$ (blue
	dash-dotted line) are shown . Also displayed are the TDSE SAE
	(red dashed line) and TDSE \textit{ab initio}  (green
	dotted line) results from \cite{2015arXiv150308966H}.}
	\label{fig:He_RABITT_angle}
\end{figure*}

Angular dependent part of the time delay in He is exhibited in
\Fref{fig:He_RABITT_angle} where we make a comparison with other
calculations reported in \cite{2015arXiv150308966H}.  Our modeling
showed that the angular-dependent part of the time delay, unlike the
energy-dependent part, is sensitive to the APT width $\tau_{\rm
APT}$. This is illustrated in the figure where we
present the two set of calculations with $\tau_{\rm APT}=2T_{\rm IR}$ and
$\tau_{\rm APT}=1.32T_{\rm IR}$ as in \cite{2015arXiv150308966H}. The
latter results are particularly close to both the SAE and \textit{ab
initio} TDSE results from \cite{2015arXiv150308966H}.


\begin{figure}[ht]
	\centering
		\includegraphics[width=0.35\textwidth,angle=-90]{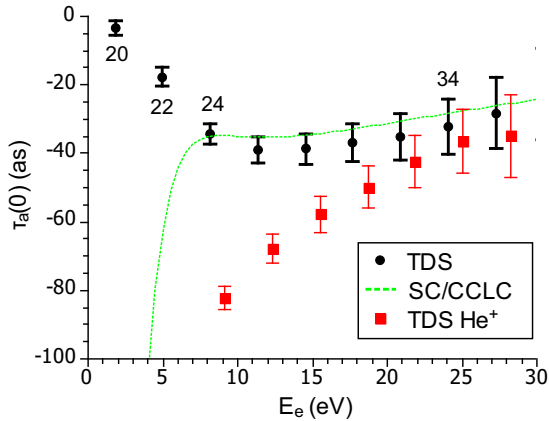}
	\caption{The time delay $\tau_a$ for H$_2^+$ as a function of
the photoelectron ejection energy $E_e$ at a fixed ejection angle
$\theta=0$. The SC/CCLC results (green dotted
line) and TDS results for He$^+$ (red squares) are also shown for comparison. Sideband indices
	made on the figure correspond to the four panels of \Fref{fig:H2p_RABITT_angle}}
	\label{fig:H2p_RABITT}
\end{figure}

\begin{figure}[ht]
	\centering
		\includegraphics[width=0.3\textwidth,angle=-90]{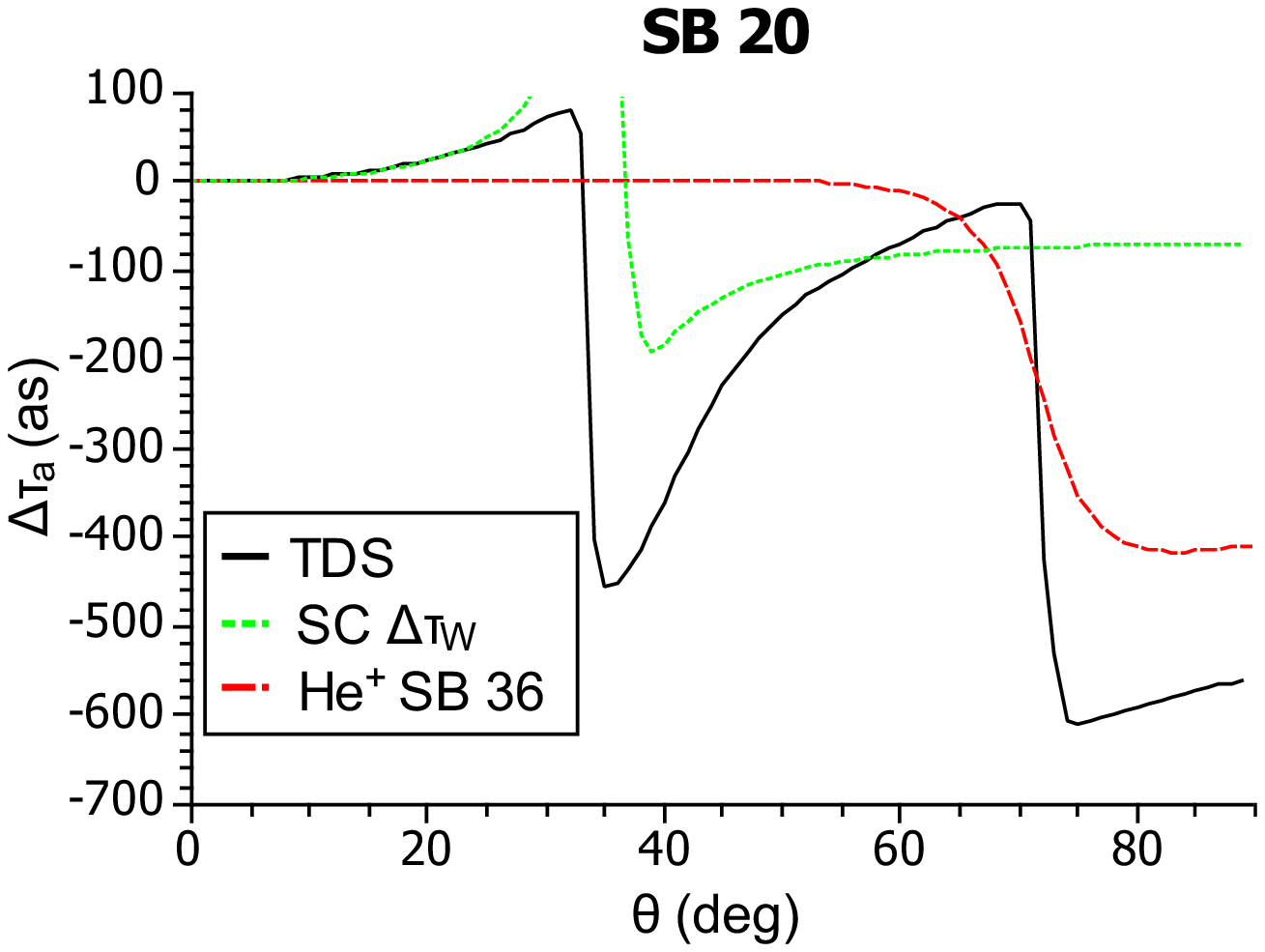}
		\includegraphics[width=0.3\textwidth,angle=-90]{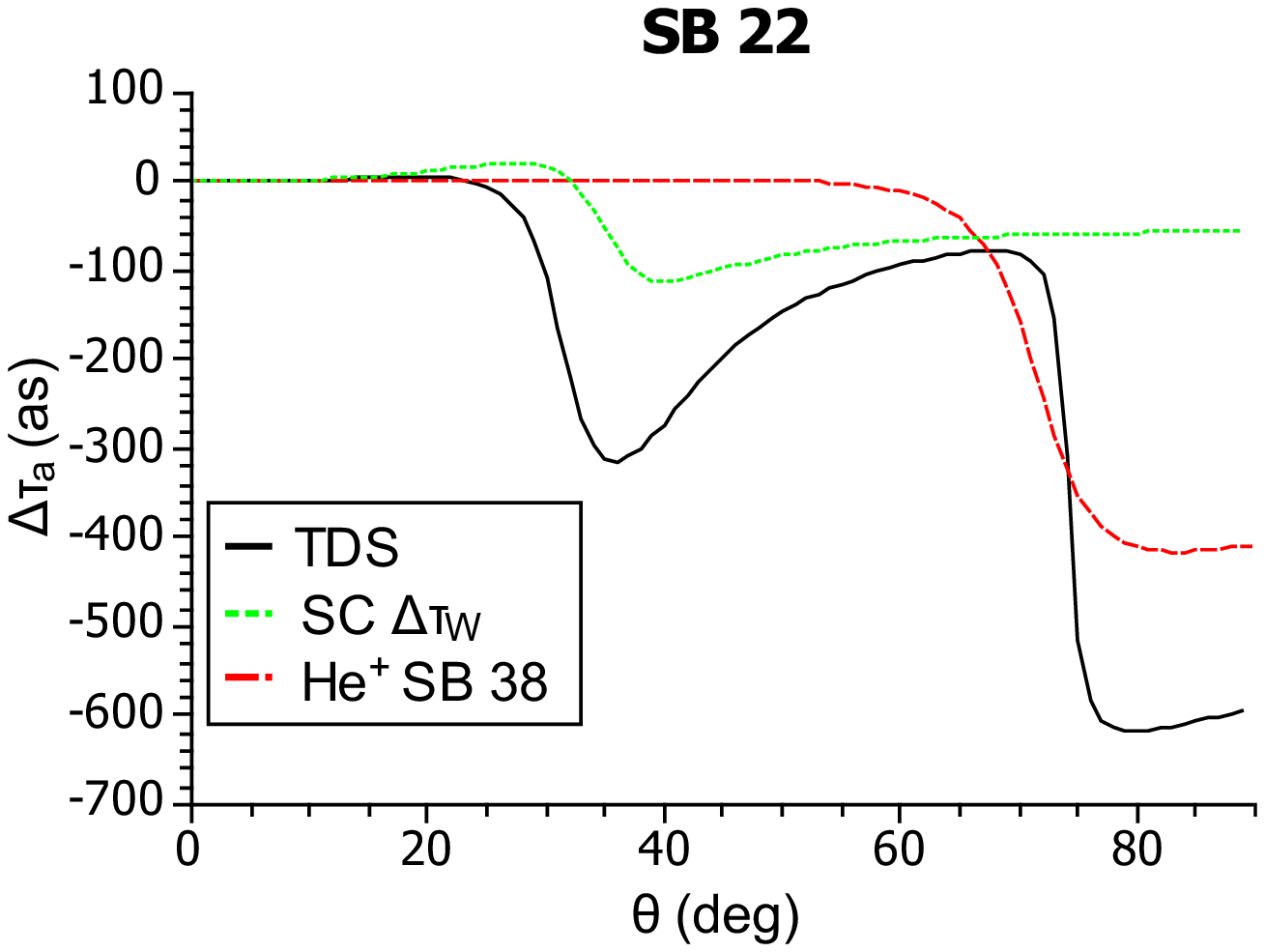}
		\includegraphics[width=0.3\textwidth,angle=-90]{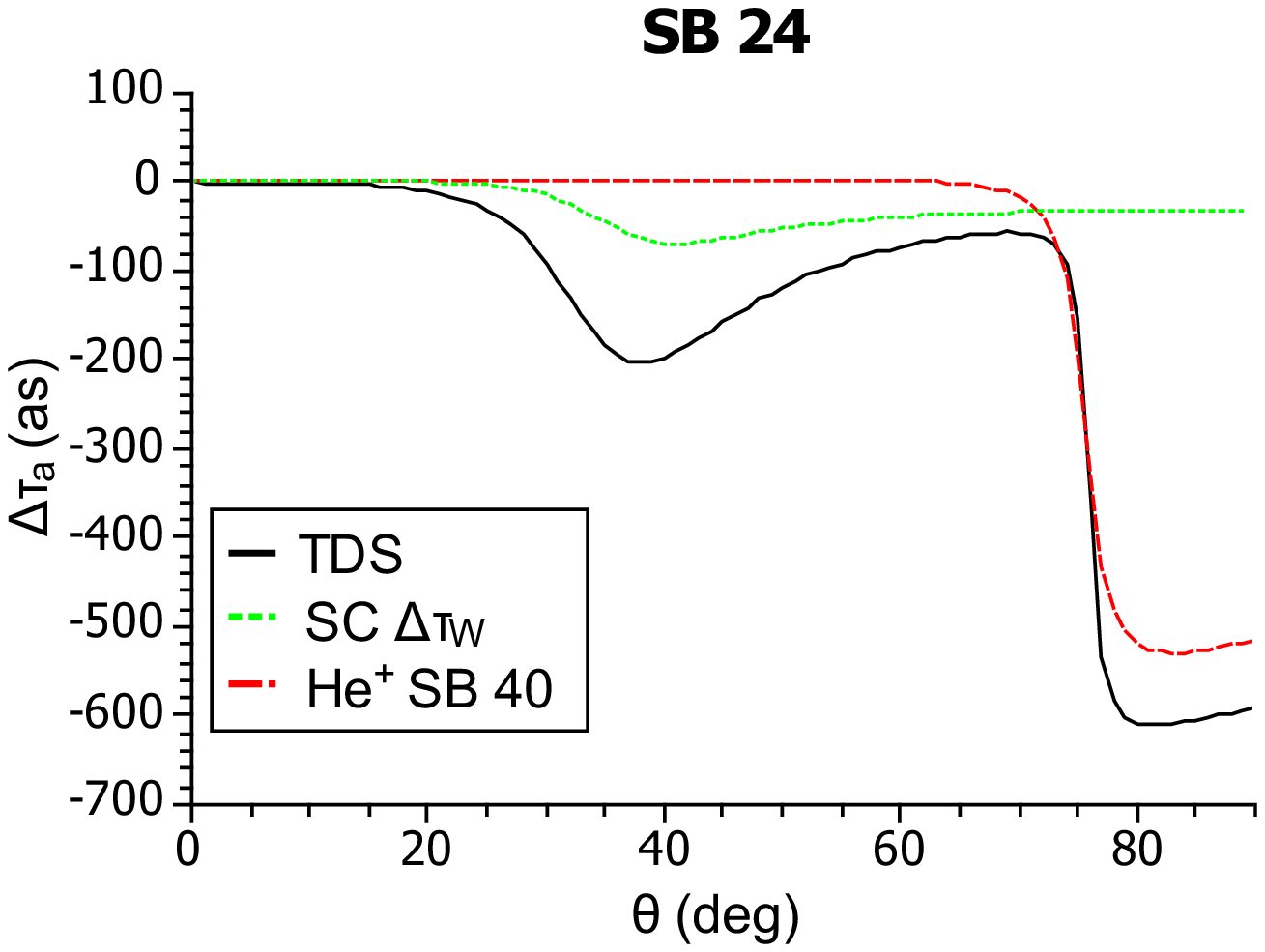}
		\includegraphics[width=0.3\textwidth,angle=-90]{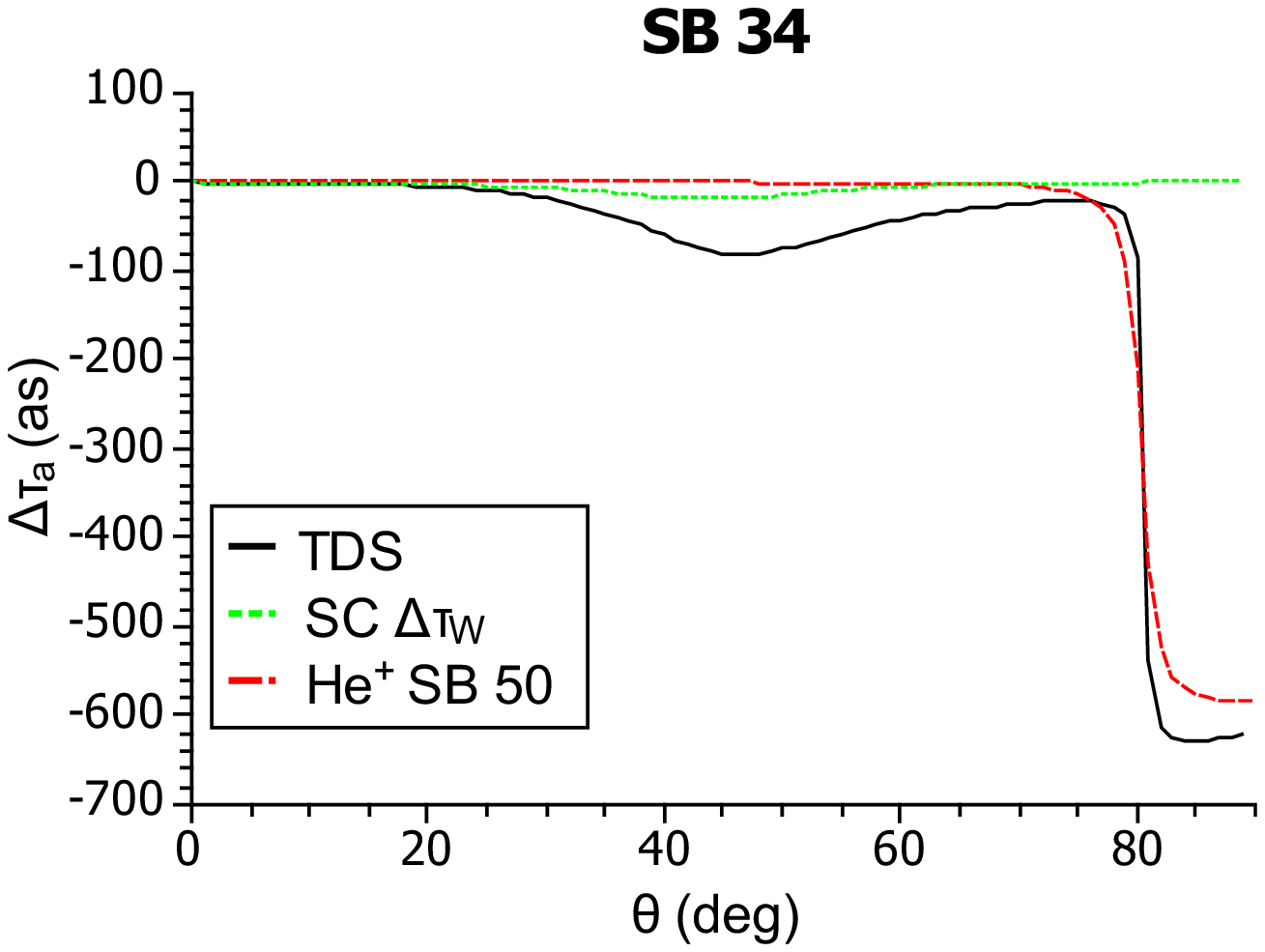}
	\caption{The angular variation of the time delay $\Delta\tau_a$ (black solid line) and Wigner's time delay $\Delta\tau_W=\tau_W(E_e,\theta)-\tau_W(E_e,0)$ (green dotted line) of H$_2^+$ for several fixed photoelectron energies $E_e$. The TDS results for He$^+$ for SBs with close energies are also shown (red dashed line).}
	\label{fig:H2p_RABITT_angle}
\end{figure}
\begin{figure}[ht]
	\centering
		\includegraphics[width=0.3\textwidth,angle=-90]{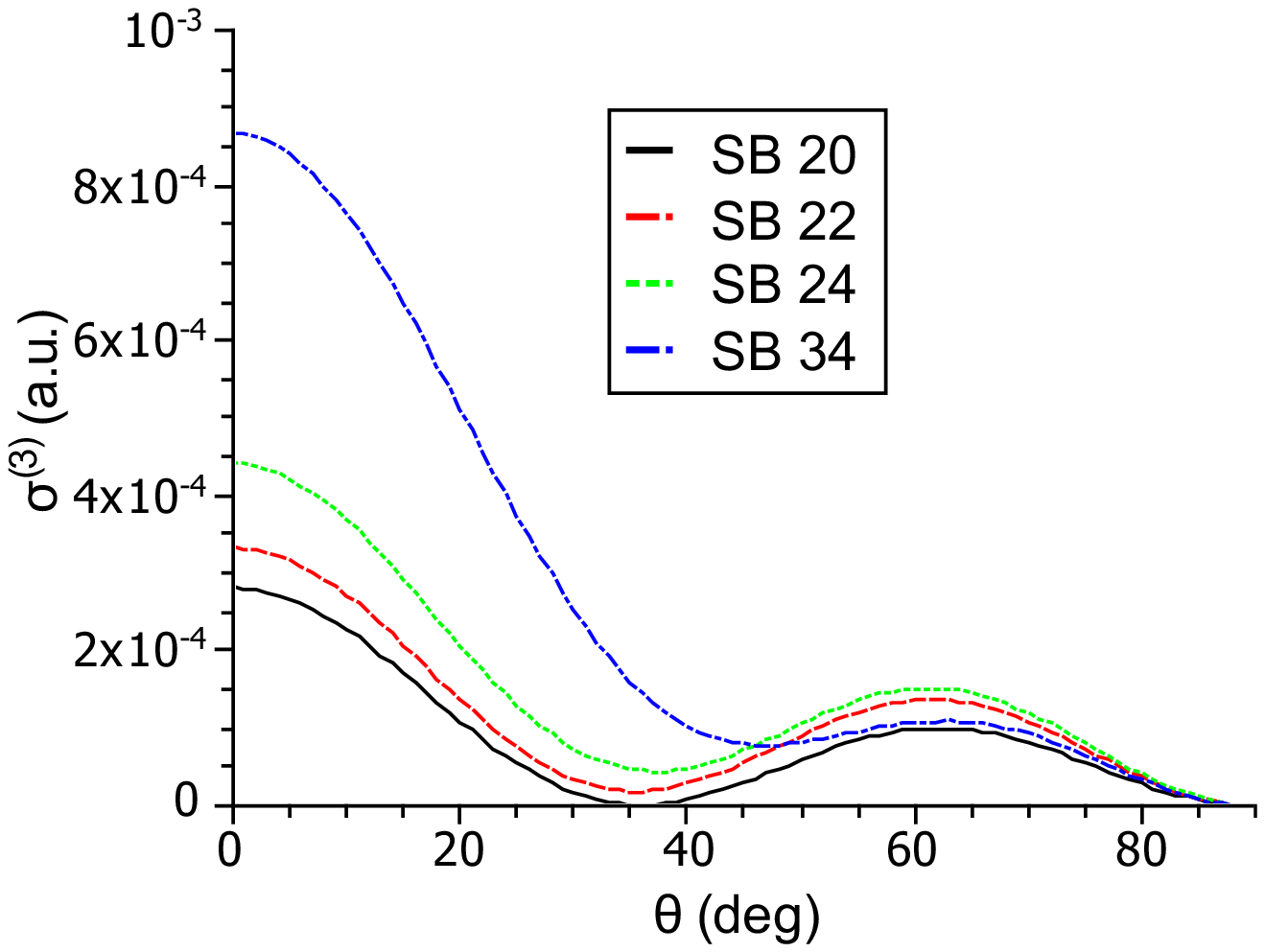}
	\caption{The triple-differential cross section $\sigma^{(3)}$ of ionization of H$_2^+$ by monochromatic XUV as function of $\theta$ for several fixed photoelectron energies $E_e$ equal SB energies on \Fref{fig:H2p_RABITT_angle}.}
	\label{fig:H2p_sigma_angle}
\end{figure}

Finally, we demonstrate the efficiency of our technique by original
calculations of the atomic time delay in the \Hp molecular ion. In
these calculations, the central frequency of the APT was set to
$\omega_{UV}=27\omega$. A polarization of the field is parallel to the molecular axis.
The energy and angular variation
of the time delay in \Hp are displayed in \Fref{fig:H2p_RABITT} and
\Fref{fig:H2p_RABITT_angle}, respectively.
Both these dependencies are very different from that of atomic H and
He.  The energy variation of $\tau_a$ with $E_e$ for H$_2^+$ is
non-monotonous.  The angular dependence of \Hp displays an additional
strong variation in the range of emission angles
$\theta=30^\circ-50^\circ$.  To visualize clearly this molecular
effect, we make a comparison of the angular dependent time delay in
\Hp with the spherically symmetric He$^+$ ion.  To account for
different ionization potentials, we carried out the He$^+$ calculation
at the central frequency $\omega_{\rm XUV}=43\omega$. It is clearly
seen that the atomic and molecular ions display the angular dependent
time delay which differs considerably not only by additional strong
angular variation but also the magnitude of the sharp drop of the time
delay near the $90^\circ$ emission angle. We note that the asymptotic
field of the ion remainder is the same in both cases. Hence should be
the same the CLC term of the atomic time delay
\eref{atomic}. Therefore the difference of the atomic time delay in
the \Hp and He$^+$ ions should be attributed largely to the Wigner
component $\tau_{\rm W}$ of the time delay.

This component is related to the monochromatic XUV photoionization
 and can be expressed via the logarithmic derivative of the
corresponding photoionization amplitude:
\be
\tau_{\rm W} = {\rm Im} 
\left [
\frac{1}{f_{\rm XUV}(\mathbf{k})}\frac{\partial f_{\rm XUV}(\mathbf{k})}{\partial E_e}
\right]. \label{WignerEq}
\ee
The angular differential XUV cross-section is expressed via the same
amplitude as
\be
\sigma^{(3)}=\frac{d^3\sigma}{dE_e d\Omega_e} = 
\frac{4\pi^2\omega_{\rm XUV}}{c} k|f_{\rm XUV}(\mathbf{k})|^2
\label{sigma}
\ee
By inspecting these two equations, we observe that the 
minimum of  the  angular differential XUV cross-section corresponds to
the maximum of the Wigner time delay. This can be indeed confirmed by
aligning \Fref{fig:H2p_RABITT_angle} with \Fref{fig:H2p_sigma_angle} where
we exhibited  the  angular differential XUV cross-section for the
corresponding sidebands. 

Ionization of \Hp by a monochromatic XUV radiation was modeled
separately by the method based on the spheroidal Coulomb (SC)
functions \cite{PhysRevA.65.062708,0953-4075-38-15-014}. With this
method, we obtained the XUV ionization amplitude and fed it to the
expression for the Wigner time delay \Eref{WignerEq} and differential
cross-section \eref{sigma}. The Wigner time delay was also estimated
by a classical approximation to CLC (CCLC) derived in
\cite{Serov2015b} for the case of $\theta=0$:
\be 
\tau_a \approx  \frac{\tau_{\rm W}+[a\ln(k/Z) +
I_c(a)]/\omega}{1-I_s(a)} \label{tS} 
\ee
with the parameter $a=Z\omega/k^3$ and functions
\ba
I_c(a)&=&-a[\ln(2/a)-1-\gamma]-\frac{3\pi}{4} a^2; \nn\\
I_s(a)&=&-\frac{\pi}{2}a+\frac{3}{2}a^2[\ln(2/a)-1/6-\gamma] \nn
\ea
Here $\gamma = 0.577$ is the Euler constant.
It is seen in \Fref{fig:H2p_RABITT} that for $E_e>8$ eV results of
SC/CCLC are rather close to those obtained from our TDS RABBITT
simulations. However, at lower energies, SC/CCLC fails. One can observe in
\Fref{fig:H2p_RABITT_angle} that the angular variation of the Wigner
time delay is qualitatively similar to variation of $\tau_a$, but a
quantitative difference is quite noticeable.  This means that the CLC
correction $\tau_{\rm CLC}$ is not a universal function that fits
\Eref{atomic} both for the He$^+$ and \Hp ions.

The interference character of the minimum in the angular differential
cross-section is revealed by its shift to the right when the photon
and photoelectron energy increase.  An
additional minimum appears at small angles when the
photoelectron energy exceeds 200~eV but this energy range is not
visualized in the figure. The relative depth of the minimum of the
angular differential cross-section increases closer to the threshold. 
Accordingly, the magnitude of the oscillation of the Wigner time delay 
and the atomic time delay grows bigger in lower side bands.
As it was demonstrated in \cite{PhysRevA.87.063414}, this the deepening
 of the minima is related to appearance of a near threshold Cooper
 minimum. This minimum has an angular character as the dipole
 component of the ionization amplitude vanishes giving way to a
 octupole component.

\section{Conclusion}

We have developed an efficient computational technique for solving the
time-dependent Schr\"odinger equation. As an illustration, we applied
this scheme to the process of two-color XUV/IR photoionization of the
molecular \Hp ion. Up to now, this process could only be described by
a simplified 2D model \cite{PhysRevA.81.062101}. We derived the energy
and angular dependent photoemission time delay and connected its
peculiarities with the photoelectron group delay (Wigner time delay)
and the Coulomb-laser coupling induced correction. The Wigner time
delay carries a strong imprint of the interference structure in the angular
resolved XUV photoionization cross-section. The Coulomb-laser coupling
correction is similar in the atomic He$^+$ and molecular \Hp ions and
is determined largely by the asymptotic part of the photoelectron wave
packet propagating in the Coulomb field of the ion remainder and the
dressing IR field.

As a further development, we will expand our technique  to
describe the photoemission time delay in  \H and other diatomic
molecules. Experimental observation of time delay in such systems has
now become possible \cite{Cattaneo2016}.

\begin{acknowledgments}
VVS acknowledges support of this work from the Russian Foundation for
Basic Research (Grant No. 14-01-00520-a). His Visiting Fellowship to
the Australian National University was supported by the Australian
Research Council Discovery Project DP120101805.
\end{acknowledgments}

\vfill\eject \np

\bibliography{references,areferences,hreferences,greferences,mypapers,../Circular-400eV/references}

\begin{thebibliography}{23}
\expandafter\ifx\csname natexlab\endcsname\relax\def\natexlab#1{#1}\fi
\expandafter\ifx\csname bibnamefont\endcsname\relax
  \def\bibnamefont#1{#1}\fi
\expandafter\ifx\csname bibfnamefont\endcsname\relax
  \def\bibfnamefont#1{#1}\fi
\expandafter\ifx\csname citenamefont\endcsname\relax
  \def\citenamefont#1{#1}\fi
\expandafter\ifx\csname url\endcsname\relax
  \def\url#1{\texttt{#1}}\fi
\expandafter\ifx\csname urlprefix\endcsname\relax\def\urlprefix{URL }\fi
\providecommand{\bibinfo}[2]{#2}
\providecommand{\eprint}[2][]{\url{#2}}

\bibitem[{\citenamefont{{Schultze {\em et~al}}}(2010)}]{M.Schultze06252010}
\bibinfo{author}{\bibfnamefont{M.}~\bibnamefont{{Schultze {\em et~al}}}},
  \emph{\bibinfo{title}{{Delay in Photoemission}}}, \bibinfo{journal}{Science}
  \textbf{\bibinfo{volume}{328}}(\bibinfo{number}{5986}), \bibinfo{pages}{1658}
  (\bibinfo{year}{2010}).

\bibitem[{\citenamefont{{Kl\"under {\em et al}
  K.}}(2011)}]{PhysRevLett.106.143002}
\bibinfo{author}{\bibfnamefont{K.}~\bibnamefont{Kl\"under}},
  \bibinfo{author}{\bibfnamefont{J.~M.}~\bibnamefont{Dahlstr\"om}},
	\bibinfo{author}{\bibfnamefont{M.}~\bibnamefont{Gisselbrecht}},
	\bibinfo{author}{\bibfnamefont{T.}~\bibnamefont{Fordell}},
	\bibinfo{author}{\bibfnamefont{M.}~\bibnamefont{Swoboda}},
  \bibinfo{author}{\bibfnamefont{D.}~\bibnamefont{Gu\'enot}},
	\bibinfo{author}{\bibfnamefont{P.}~\bibnamefont{Johnsson}},
	\bibinfo{author}{\bibfnamefont{J.}~\bibnamefont{Caillat}},
  \bibinfo{author}{\bibfnamefont{J.}~\bibnamefont{Mauritsson}},
  \bibinfo{author}{\bibfnamefont{A.}~\bibnamefont{Maquet}},
  \bibinfo{author}{\bibfnamefont{R.}~\bibnamefont{Ta\"ieb}}, \bibnamefont{and}
	\bibinfo{author}{\bibfnamefont{A.} \bibnamefont{L’Huillier}},
  \emph{\bibinfo{title}{Probing single-photon ionization on the attosecond time
  scale}}, \bibinfo{journal}{Phys. Rev. Lett.}
  \textbf{\bibinfo{volume}{106}}(\bibinfo{number}{14}), \bibinfo{pages}{143002}
  (\bibinfo{year}{2011}).

\bibitem[{\citenamefont{Pazourek et~al.}(2015)\citenamefont{Pazourek, Nagele,
  and Burgd\"orfer}}]{RevModPhys.87.765}
\bibinfo{author}{\bibfnamefont{R.}~\bibnamefont{Pazourek}},
  \bibinfo{author}{\bibfnamefont{S.}~\bibnamefont{Nagele}}, \bibnamefont{and}
  \bibinfo{author}{\bibfnamefont{J.}~\bibnamefont{Burgd\"orfer}},
  \emph{\bibinfo{title}{Attosecond chronoscopy of photoemission}},
  \bibinfo{journal}{Rev. Mod. Phys.} \textbf{\bibinfo{volume}{87}},
  \bibinfo{pages}{765} (\bibinfo{year}{2015}).

\bibitem[{\citenamefont{W\"atzel et~al.}(2015)\citenamefont{W\"atzel,
  Moskalenko, Pavlyukh, and Berakdar}}]{0953-4075-48-2-025602}
\bibinfo{author}{\bibfnamefont{J.}~\bibnamefont{W\"atzel}},
  \bibinfo{author}{\bibfnamefont{A.~S.} \bibnamefont{Moskalenko}},
  \bibinfo{author}{\bibfnamefont{Y.}~\bibnamefont{Pavlyukh}}, \bibnamefont{and}
  \bibinfo{author}{\bibfnamefont{J.}~\bibnamefont{Berakdar}},
  \emph{\bibinfo{title}{Angular resolved time delay in photoemission}},
  \bibinfo{journal}{J. Phys. B}
  \textbf{\bibinfo{volume}{48}}(\bibinfo{number}{2}), \bibinfo{pages}{025602}
  (\bibinfo{year}{2015}).

\bibitem[{\citenamefont{Dahlstr\"om and Lindroth}(2014)}]{Dahlstrom2014}
\bibinfo{author}{\bibfnamefont{J.~M.} \bibnamefont{Dahlstr\"om}}
  \bibnamefont{and} \bibinfo{author}{\bibfnamefont{E.}~\bibnamefont{Lindroth}},
  \emph{\bibinfo{title}{Study of attosecond delays using perturbation diagrams
  and exterior complex scaling}}, \bibinfo{journal}{J. Phys. B}
  \textbf{\bibinfo{volume}{47}}(\bibinfo{number}{12}), \bibinfo{pages}{124012}
  (\bibinfo{year}{2014}).

\bibitem[{\citenamefont{{Heuser} et~al.}(2016)\citenamefont{{Heuser},
  {Jim{\'e}nez Gal{\'a}n}, {Cirelli}, {Sabbar}, {Boge}, {Lucchini}, {Gallmann},
  {Ivanov}, {Kheifets}, {Dahlstr{\"o}m} et~al.}}]{2015arXiv150308966H}
\bibinfo{author}{\bibfnamefont{S.}~\bibnamefont{{Heuser}}},
  \bibinfo{author}{\bibfnamefont{{\'A}.}~\bibnamefont{{Jim{\'e}nez
  Gal{\'a}n}}}, \bibinfo{author}{\bibfnamefont{C.}~\bibnamefont{{Cirelli}}},
  \bibinfo{author}{\bibfnamefont{M.}~\bibnamefont{{Sabbar}}},
  \bibinfo{author}{\bibfnamefont{R.}~\bibnamefont{{Boge}}},
  \bibinfo{author}{\bibfnamefont{M.}~\bibnamefont{{Lucchini}}},
  \bibinfo{author}{\bibfnamefont{L.}~\bibnamefont{{Gallmann}}},
  \bibinfo{author}{\bibfnamefont{I.}~\bibnamefont{{Ivanov}}},
  \bibinfo{author}{\bibfnamefont{A.~S.} \bibnamefont{{Kheifets}}},
  \bibinfo{author}{\bibfnamefont{J.~M.} \bibnamefont{{Dahlstr{\"o}m}}},
  \bibnamefont{et~al.}, \emph{\bibinfo{title}{Time delay anisotropy in
  photoelectron emission from the isotropic ground state of helium}},
  \bibinfo{journal}{ArXiv e-prints 1503.08966, Nat. Comm. submitted}
  (\bibinfo{year}{2016}).

\bibitem[{\citenamefont{Serov et~al.}(2013)\citenamefont{Serov, Derbov, and
  Sergeeva}}]{PhysRevA.87.063414}
\bibinfo{author}{\bibfnamefont{V.~V.} \bibnamefont{Serov}},
  \bibinfo{author}{\bibfnamefont{V.~L.} \bibnamefont{Derbov}},
  \bibnamefont{and} \bibinfo{author}{\bibfnamefont{T.~A.}
  \bibnamefont{Sergeeva}}, \emph{\bibinfo{title}{Interpretation of time delay
  in the ionization of two-center systems}}, \bibinfo{journal}{Phys. Rev. A}
  \textbf{\bibinfo{volume}{87}}, \bibinfo{pages}{063414}
  (\bibinfo{year}{2013}).

\bibitem[{\citenamefont{Dahlstr\"om et~al.}(2012)\citenamefont{Dahlstr\"om,
  Gu\'enot, Kl\"under, Gisselbrecht, Mauritsson, Huillier, Maquet, and
  Ta\"ieb}}]{Dahlstrom2012}
\bibinfo{author}{\bibfnamefont{J.}~\bibnamefont{Dahlstr\"om}},
  \bibinfo{author}{\bibfnamefont{D.}~\bibnamefont{Gu\'enot}},
  \bibinfo{author}{\bibfnamefont{K.}~\bibnamefont{Kl\"under}},
  \bibinfo{author}{\bibfnamefont{M.}~\bibnamefont{Gisselbrecht}},
  \bibinfo{author}{\bibfnamefont{J.}~\bibnamefont{Mauritsson}},
  \bibinfo{author}{\bibfnamefont{A.~L.} \bibnamefont{Huillier}},
  \bibinfo{author}{\bibfnamefont{A.}~\bibnamefont{Maquet}}, \bibnamefont{and}
  \bibinfo{author}{\bibfnamefont{R.}~\bibnamefont{Ta\"ieb}},
  \emph{\bibinfo{title}{Theory of attosecond delays in laser-assisted
  photoionization}}, \bibinfo{journal}{Chem. Phys.}
  \textbf{\bibinfo{volume}{414}}, \bibinfo{pages}{53} (\bibinfo{year}{2012}).

\bibitem[{\citenamefont{Ivanov and Kheifets}(2013)}]{Whelan2013}
\bibinfo{author}{\bibfnamefont{I.}~\bibnamefont{Ivanov}} \bibnamefont{and}
  \bibinfo{author}{\bibfnamefont{A.}~\bibnamefont{Kheifets}},
  \emph{\bibinfo{title}{Fragmentation Processes}}
  (\bibinfo{publisher}{Cambridge University Press}, \bibinfo{year}{2013}),
  chap. \bibinfo{chapter}{Atoms with one and two active electrons in strong
  laser fields}, Topics in Atomic and Molecular Physics.

\bibitem[{\citenamefont{Jim\'enez-Gal\'an
  et~al.}(2014)\citenamefont{Jim\'enez-Gal\'an, Argenti, and
  Mart\'{\i}n}}]{PhysRevLett.113.263001}
\bibinfo{author}{\bibfnamefont{A.}~\bibnamefont{Jim\'enez-Gal\'an}},
  \bibinfo{author}{\bibfnamefont{L.}~\bibnamefont{Argenti}}, \bibnamefont{and}
  \bibinfo{author}{\bibfnamefont{F.}~\bibnamefont{Mart\'{\i}n}},
  \emph{\bibinfo{title}{Modulation of attosecond beating in resonant two-photon
  ionization}}, \bibinfo{journal}{Phys. Rev. Lett.}
  \textbf{\bibinfo{volume}{113}}, \bibinfo{pages}{263001}
  (\bibinfo{year}{2014}).

\bibitem[{\citenamefont{Serov et~al.}(2007)\citenamefont{Serov, Derbov,
  Joulakian, and Vinitsky}}]{PhysRevA.75.012715}
\bibinfo{author}{\bibfnamefont{V.~V.} \bibnamefont{Serov}},
  \bibinfo{author}{\bibfnamefont{V.~L.} \bibnamefont{Derbov}},
  \bibinfo{author}{\bibfnamefont{B.~B.} \bibnamefont{Joulakian}},
  \bibnamefont{and} \bibinfo{author}{\bibfnamefont{S.~I.}
  \bibnamefont{Vinitsky}}, \emph{\bibinfo{title}{Wave-packet-evolution approach
  for single and double ionization of two-electron systems by fast electrons}},
  \bibinfo{journal}{Phys. Rev. A} \textbf{\bibinfo{volume}{75}},
  \bibinfo{pages}{012715} (\bibinfo{year}{2007}).

\bibitem[{\citenamefont{{Serov}}(2015)}]{2015arXiv150907115S}
\bibinfo{author}{\bibfnamefont{V.~V.} \bibnamefont{{Serov}}},
  \emph{\bibinfo{title}{{Orthogonal fast spherical Bessel transform on uniform
  grid}}}, \bibinfo{journal}{ArXiv e-prints}  (\bibinfo{year}{2015}),
  \eprint{1509.07115}.

\bibitem[{\citenamefont{{Akoury {\em et al}}}(2007)}]{Akoury09112007}
\bibinfo{author}{\bibfnamefont{D.}~\bibnamefont{{Akoury {\em et al}}}},
  \emph{\bibinfo{title}{The simplest double slit: Interference and entanglement
  in double photoionization of {H$_2$}}}, \bibinfo{journal}{Science}
  \textbf{\bibinfo{volume}{318}}(\bibinfo{number}{5852}), \bibinfo{pages}{949}
  (\bibinfo{year}{2007}).

\bibitem[{\citenamefont{Chelkowski and Bandrauk}(2010)}]{PhysRevA.81.062101}
\bibinfo{author}{\bibfnamefont{S.}~\bibnamefont{Chelkowski}} \bibnamefont{and}
  \bibinfo{author}{\bibfnamefont{A.~D.} \bibnamefont{Bandrauk}},
  \emph{\bibinfo{title}{Visualizing electron delocalization, electron-proton
  correlations, and the {Einstein}-{Podolsky}-{Rosen} paradox during the
  photodissociation of a diatomic molecule using two ultrashort laser pulses}},
  \bibinfo{journal}{Phys. Rev. A} \textbf{\bibinfo{volume}{81}},
  \bibinfo{pages}{062101} (\bibinfo{year}{2010}).

\bibitem[{\citenamefont{Wigner}(1955)}]{PhysRev.98.145}
\bibinfo{author}{\bibfnamefont{E.~P.} \bibnamefont{Wigner}},
  \emph{\bibinfo{title}{Lower limit for the energy derivative of the scattering
  phase shift}}, \bibinfo{journal}{Phys. Rev.}
  \textbf{\bibinfo{volume}{98}}(\bibinfo{number}{1}), \bibinfo{pages}{145}
  (\bibinfo{year}{1955}).

\bibitem[{\citenamefont{Nagele et~al.}(2011)\citenamefont{Nagele, Pazourek,
  Feist, Doblhoff-Dier, Lemell, T\"ok\'esi, and
  Burgd\"orfer}}]{0953-4075-44-8-081001}
\bibinfo{author}{\bibfnamefont{S.}~\bibnamefont{Nagele}},
  \bibinfo{author}{\bibfnamefont{R.}~\bibnamefont{Pazourek}},
  \bibinfo{author}{\bibfnamefont{J.}~\bibnamefont{Feist}},
  \bibinfo{author}{\bibfnamefont{K.}~\bibnamefont{Doblhoff-Dier}},
  \bibinfo{author}{\bibfnamefont{C.}~\bibnamefont{Lemell}},
  \bibinfo{author}{\bibfnamefont{K.}~\bibnamefont{T\"ok\'esi}},
  \bibnamefont{and}
  \bibinfo{author}{\bibfnamefont{J.}~\bibnamefont{Burgd\"orfer}},
  \emph{\bibinfo{title}{Time-resolved photoemission by attosecond streaking:
  extraction of time information}}, \bibinfo{journal}{J.~Phys.~B}
  \textbf{\bibinfo{volume}{44}}(\bibinfo{number}{8}), \bibinfo{pages}{081001}
  (\bibinfo{year}{2011}).

\bibitem[{\citenamefont{Paul et~al.}(2001)\citenamefont{Paul, Toma, Breger,
  Mullot, Aug\'e, Balcou, Muller, and Agostini}}]{Paul01062001}
\bibinfo{author}{\bibfnamefont{P.~M.} \bibnamefont{Paul}},
  \bibinfo{author}{\bibfnamefont{E.~S.} \bibnamefont{Toma}},
  \bibinfo{author}{\bibfnamefont{P.}~\bibnamefont{Breger}},
  \bibinfo{author}{\bibfnamefont{G.}~\bibnamefont{Mullot}},
  \bibinfo{author}{\bibfnamefont{F.}~\bibnamefont{Aug\'e}},
  \bibinfo{author}{\bibfnamefont{P.}~\bibnamefont{Balcou}},
  \bibinfo{author}{\bibfnamefont{H.~G.} \bibnamefont{Muller}},
  \bibnamefont{and} \bibinfo{author}{\bibfnamefont{P.}~\bibnamefont{Agostini}},
  \emph{\bibinfo{title}{Observation of a train of attosecond pulses from high
  harmonic generation}}, \bibinfo{journal}{Science}
  \textbf{\bibinfo{volume}{292}}(\bibinfo{number}{5522}), \bibinfo{pages}{1689}
  (\bibinfo{year}{2001}).

\bibitem[{\citenamefont{Sarsa et~al.}(2004)\citenamefont{Sarsa, G\'{a}lvez, and
  Buendia}}]{Sarsa2004163}
\bibinfo{author}{\bibfnamefont{A.}~\bibnamefont{Sarsa}},
  \bibinfo{author}{\bibfnamefont{F.~J.} \bibnamefont{G\'{a}lvez}},
  \bibnamefont{and} \bibinfo{author}{\bibfnamefont{E.}~\bibnamefont{Buendia}},
  \emph{\bibinfo{title}{Parameterized optimized effective potential for the
  ground state of the atoms {He} through {Xe}}}, \bibinfo{journal}{Atomic Data
  and Nuclear Data Tables} \textbf{\bibinfo{volume}{88}}(\bibinfo{number}{1}),
  \bibinfo{pages}{163 } (\bibinfo{year}{2004}).

\bibitem[{\citenamefont{Serov et~al.}(2002)\citenamefont{Serov, Joulakian,
  Pavlov, Puzynin, and Vinitsky}}]{PhysRevA.65.062708}
\bibinfo{author}{\bibfnamefont{V.~V.} \bibnamefont{Serov}},
  \bibinfo{author}{\bibfnamefont{B.~B.} \bibnamefont{Joulakian}},
  \bibinfo{author}{\bibfnamefont{D.~V.} \bibnamefont{Pavlov}},
  \bibinfo{author}{\bibfnamefont{I.~V.} \bibnamefont{Puzynin}},
  \bibnamefont{and} \bibinfo{author}{\bibfnamefont{S.~I.}
  \bibnamefont{Vinitsky}}, \emph{\bibinfo{title}{$(e,2e)$ ionization of
  {H}$_2^+$ by fast electron impact: Application of the exact nonrelativistic
  two-center continuum wave}}, \bibinfo{journal}{Phys. Rev. A}
  \textbf{\bibinfo{volume}{65}}, \bibinfo{pages}{062708}
  (\bibinfo{year}{2002}).

\bibitem[{\citenamefont{Serov et~al.}(2005)\citenamefont{Serov, Joulakian,
  Derbov, and Vinitsky}}]{0953-4075-38-15-014}
\bibinfo{author}{\bibfnamefont{V.~V.} \bibnamefont{Serov}},
  \bibinfo{author}{\bibfnamefont{B.~B.} \bibnamefont{Joulakian}},
  \bibinfo{author}{\bibfnamefont{V.~L.} \bibnamefont{Derbov}},
  \bibnamefont{and} \bibinfo{author}{\bibfnamefont{S.~I.}
  \bibnamefont{Vinitsky}}, \emph{\bibinfo{title}{Ionization excitation of
  diatomic systems having two active electrons by fast electron impact: a probe
  to electron correlation}}, \bibinfo{journal}{J.~Phys.~B}
  \textbf{\bibinfo{volume}{38}}(\bibinfo{number}{15}), \bibinfo{pages}{2765}
  (\bibinfo{year}{2005}).

\bibitem[{\citenamefont{Serov et~al.}(2015)\citenamefont{Serov, Derbov, and
  Sergeeva}}]{Serov2015b}
\bibinfo{author}{\bibfnamefont{V.~V.} \bibnamefont{Serov}},
  \bibinfo{author}{\bibfnamefont{V.~L.} \bibnamefont{Derbov}},
  \bibnamefont{and} \bibinfo{author}{\bibfnamefont{T.~A.}
  \bibnamefont{Sergeeva}}, \emph{\bibinfo{title}{Interpretation of the time
  delay in the ionization of {Coulomb} systems by attosecond laser pulses}}, in
  \emph{\bibinfo{booktitle}{Advanced Lasers}} (\bibinfo{publisher}{Springer},
  \bibinfo{address}{Berlin}, \bibinfo{year}{2015}), vol. \bibinfo{volume}{193}
  of \emph{\bibinfo{series}{Springer Series in Optical Sciences}}, pp.
  \bibinfo{pages}{213--230}.

\bibitem[{\citenamefont{Vos et~al.}(2016)\citenamefont{Vos, Cattaneo, Heuser,
  Lucchini, Cirelli, and Keller}}]{Cattaneo2016}
\bibinfo{author}{\bibfnamefont{J.}~\bibnamefont{Vos}},
  \bibinfo{author}{\bibfnamefont{L.}~\bibnamefont{Cattaneo}},
  \bibinfo{author}{\bibfnamefont{S.}~\bibnamefont{Heuser}},
  \bibinfo{author}{\bibfnamefont{M.}~\bibnamefont{Lucchini}},
  \bibinfo{author}{\bibfnamefont{C.}~\bibnamefont{Cirelli}}, \bibnamefont{and}
  \bibinfo{author}{\bibfnamefont{U.}~\bibnamefont{Keller}},
  \emph{\bibinfo{title}{Asymmetric {Wigner} time delay in {CO}
  photoionization}}, in \emph{\bibinfo{booktitle}{12th European Conference on
  Atoms, Molecules and Photons}} (\bibinfo{address}{Frankfurt},
  \bibinfo{year}{2016}).

\end{thebibliography}

\end{document}